\documentclass[onecolumn]{emulateapj}

\usepackage{epsfig}
\usepackage{graphicx}
\usepackage{epsf}

\usepackage{natbib}

\def\apj{{\em Ap.~J.}~}
\def\aa{{\em Astr.~Ap.}~}
\def\aap{{\em Astr.~Ap.}~}
\def\prl{{\em Phys. Rev. Letts.}~}
\def\prd{{\em Phys. Rev. D}~}
\def\sgreat{\lower2pt\hbox{$\buildrel {\scriptstyle >}
   \over {\scriptstyle\sim}$}}

\begin{document}
\title{Multi-Dimensional Explorations in Supernova Theory}
\author{Adam Burrows\altaffilmark{1}, Luc Dessart\altaffilmark{1}, 
Christian D. Ott\altaffilmark{2}, \& Eli Livne\altaffilmark{3}}

\altaffiltext{1}{Department of Astronomy and Steward Observatory,
                 The University of Arizona, Tucson, AZ \ 85721;
                 burrows@as.arizona.edu,luc@as.arizona.edu}
\altaffiltext{2}{Max-Planck-Institut f\"{u}r Gravitationsphysik,
Albert-Einstein-Institut, Golm/Potsdam, Germany; cott@aei.mpg.de}
\altaffiltext{3}{Racah Institute of Physics, The Hebrew University,
Jerusalem, Israel; eli@frodo.fiz.huji.ac.il}

\begin{abstract}

In this paper, we bring together various of our published and 
unpublished findings from our recent 2D multi-group, flux-limited
radiation hydrodynamic simulations of the collapse and explosion of the cores
of massive stars. Aided by 2D and 3D graphical renditions, we motivate
the acoustic mechanism of core-collapse supernova explosions and explain, as best we currently can,
the phases and phenomena that attend this mechanism.   Two major foci of our presentation are the outer
shock instability and the inner core g-mode oscillations.  The former sets the stage for 
the latter, which damp by the generation of sound.  This sound propagates outward to energize the explosion
and is relevant only if the core has not exploded earlier by some other means.  Hence,
it is a more delayed mechanism than the traditional neutrino mechanism that has been studied 
for the last twenty years since it was championed by Bethe and Wilson.  We discuss
protoneutron star convection, accretion-induced-collapse, gravitational wave emissions,
pulsar kicks, the angular anisotropy of the neutrino emissions, a subset of numerical issues,
and a new code we are designing that should supercede our current 
supernova code VULCAN/2D.  Whatever ideas last from this current generation of
numerical results, and whatever the eventual mechanism(s), we conclude that the breaking 
of spherical symmetry will survive as one of the crucial keys to the supernova puzzle.

\end{abstract}


\section{Introduction}
\label{intro}

Over the past forty-five years, the theory of core-collapse supernova explosions 
and attendant phenomena has witnessed evolution, elaboration, clarification, and 
much speculation.  Many ideas have come and gone, but in the process the physics, initial conditions,
and numerical sophistication have all greatly improved.  Such has been the progress
that there are some who believe that the physical inputs are in hand and it is only the numerical 
tools that need sharpening and development.  That may be, but even if this were the case
such is the complexity of the three-dimensional radiation hydrodynamics (magneto-hydrodynamics?)
which Nature integrates effortlessly and in real time that astrophysicists are
still debating the primary agents of explosion.  The major drivers that remain
are neutrinos, acoustic power, and magnetic stresses. The neutrino heating 
mechanism is still the favorite, but both acoustic power due to the excitation
of inner core pulsations and magnetic hoop stresses generated in a rapidly differentially 
rotating protoneutron star core are contenders. It may be a mix of all three.   

The central issues of supernova explosion theory are 1) the mechanism of explosion; 2)  
the nucleosynthetic yields, including $^{56}$Ni, the rp-process, and the r-process; 3) the blast
morphology (and the associated polarization of the emergent light); 4) the origin of pulsar kicks; 5)
the origin of pulsar spins; 6) the generation and distribution of pulsar, AXP, and magnetar B-fields; 7)
black hole formation; 8) the systematics of all these things with progenitor, its mass, metallicity, 
and rotational profile; and 9) the connection with gamma-ray bursts.  We are still 
a long way from achieving this program, but there are new ideas and numerical results 
that are beginning to resemble, however imperfectly, a suggestive idealization of Nature.   
In this paper, we review and synthesize our emerging perspectives on some of these issues,
with a bias towards our own work and thoughts.  To do so, we have bought together in one
place various conclusions that can be found scattered in many of our recently published 
works, as well as new results and graphics that add to the overall analysis and 
discussion. What we conclude is that, whatever the central mechanism of explosion, 
the breaking of spherical symmetry, multi-dimensional effects, and instabilities 
are keys to the supernova puzzle.  This is the theme, if there is one, of this monograph.
Not only do pulsar kicks hint at this, but all of the most sophisticated simulations recently 
undertaken point in this direction, whether the mechanism is via neutrinos or sound.   
What survives of this current generation of simulation results and ideas will depend 
upon the next generation of calculations, and the addition of adequate 3D simulations 
and analysis.  Hence, we provide here but a snapshot of the moving theoretical terrain.

\section{The Neutrino Mechanism?}
\label{neutrino_mechanism}

\begin{figure*}
\epsfig{file=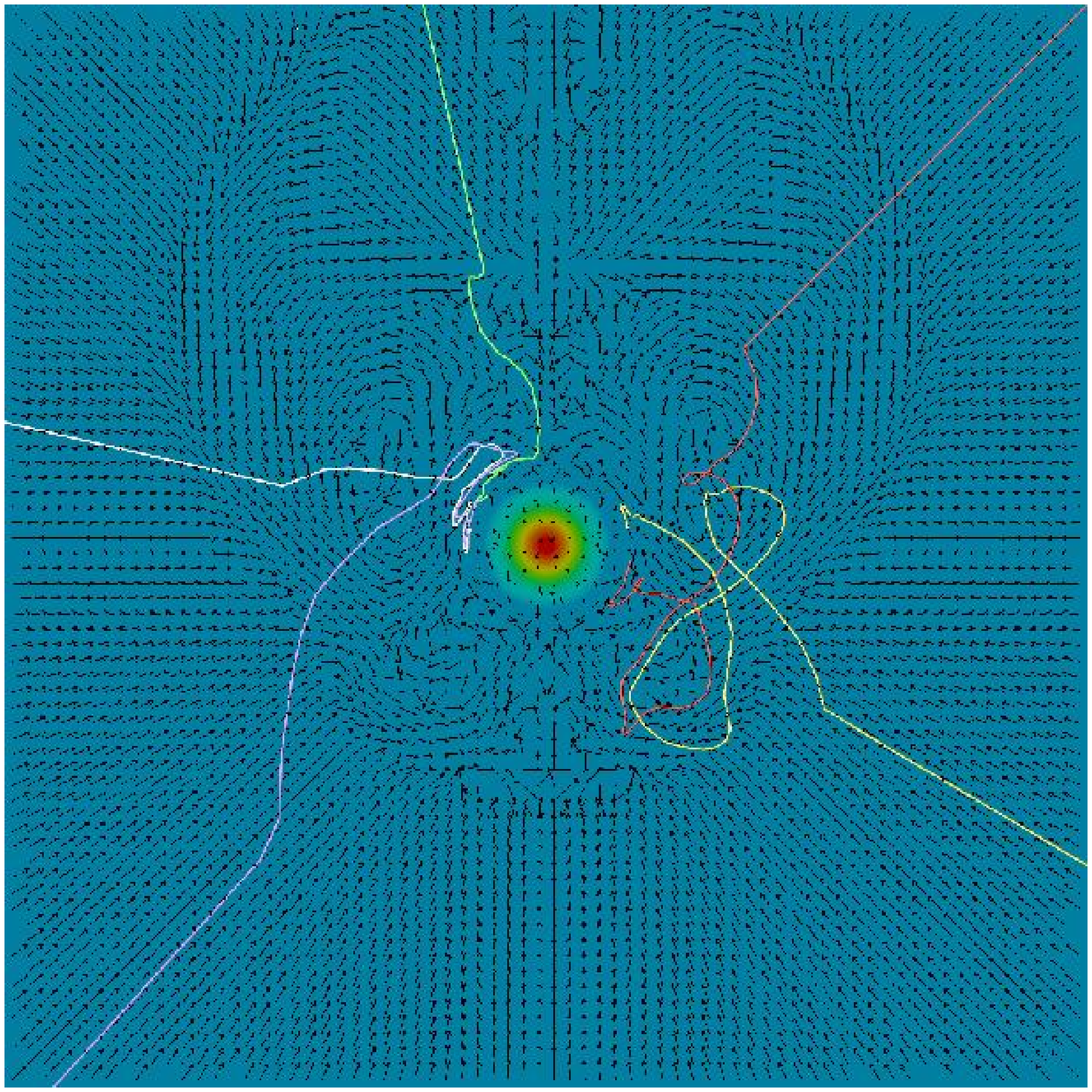, width=0.5\textwidth}
\epsfig{file=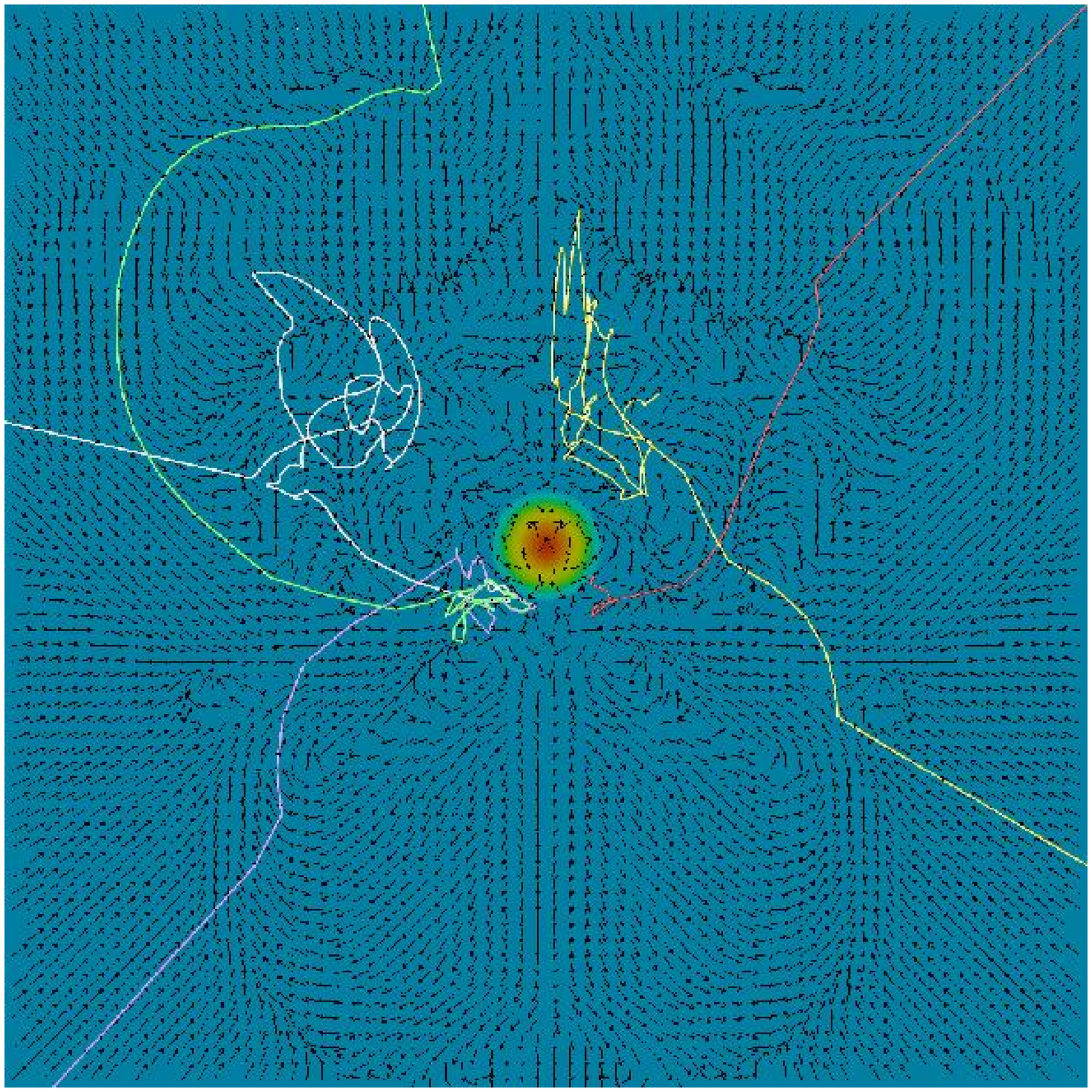, width=0.5\textwidth}
\caption{
Two renderings of five representative streak lines showing the paths of infalling parcels of matter
during a period of $\sim$100 milliseconds.  The starting points were two different random times
during the SASI phase and before explosion. Clearly seen is the encounter with the shock
and the subsequent effects of the turbulent/convective post-shock environment.
The velocity vector fields in the backgrounds are for the ``final" times.
In 1D, these would be straight lines along a radius.  In this representative 2D
simulation, the matter is shown to dwell a bit longer in the outer shocked region
before it settles onto the inner core.
}
\label{aburrows_streak}
\end{figure*}

The neutrino-heating mechanism, in which a stalled bounce shock is reenergized
by neutrino energy deposition after a slight delay, perhaps aided by overturning
instabilities in this ``gain region," has been the working hypothesis of supernova theorists for
the last twenty years \cite{bethe,buras2006,buras2,lieben2001}. Past calculations in support of
this mechanism, or variations on its theme, include those by Wilson \& Mayle \cite{wm88,wm93},
Mayle \& Wilson \cite{mayle}, Herant et al. \cite{herant}, Burrows, Hayes, \& Fryxell \cite{bhf},
Janka \& M\"uller \cite{muller96}, and Fryer \& Warren \cite{fryer2002,fryer2004}.  Convection was thought
to facilitate explosion by increasing the dwell time in the gain region of 
parcels of matter accreting through the shock, and thereby the efficiency  
of neutrino heating.  In 1D, such parcels move quickly through the gain region, 
achieve the cooling region, and settle upon the core before the explosion
condition is met. Figure \ref{aburrows_streak} depicts examples of the circuitous routes taken
by accreting elements after they penetrate through the shock.
%
However, recent calculations employing careful
neutrino physics and numerics suggest that the neutrino mechanism,
when it succeeds, may at best be marginal.
Kitaura et al. \cite{kitaura} follow in spherical symmetry the
compact 1.38-M$_{\odot}$ O-Ne-Mg core of the 8.8-M$_{\odot}$ model
of Nomoto \& Hashimoto \cite{nomoto}, with a very tenuous outer envelope, and
obtain a delayed neutrino-driven explosion. 
However, the explosion energy is only $\sgreat$10$^{50}$ ergs.  
For the slightly more massive 11.2-M$_{\odot}$ progenitor of Woosley, 
Heger, \& Weaver \cite{woosley02} (WHW02), Buras et al. \cite{buras2} witness the onset, 
not in 1D, but in 2D, of a neutrino-driven dipolar explosion aided by the 
the advective-acoustic Standing-Accretion-Shock-Instability (SASI) 
\cite{blond03,blond06,fogt00,fog01,fog02,fog5a,buras2006,buras2,bur06,bur06b,fog06}.   
They also infer a very weak explosion energy, this time near 10$^{49}$ ergs, not correcting for
neutrino driving subsequent to the early termination of their calculation or for the
binding energy of the outer envelope.  Buras et al. \cite{buras2} focus on the importance
of calculating over the full 180$^{\circ}$ 2D domain so as not to suppress the $\ell = 1$ mode
of the SASI, since they do not obtain even a weak explosion when
constraining the computational domain to 90$^{\circ}$.  Importantly, the breaking of spherical symmetry
is a key to the explosion they see, an emerging theme that we endorse
for the majority of core-collapse explosions.  However, Buras et al. \cite{buras2} don't
as yet see any indications that more massive progenitors explode by the neutrino mechanism,
even when aided by the SASI and convection.

\section{The Acoustic Mechanism?}
\label{acoustic_mechanism}

\begin{figure*}
\epsfig{file=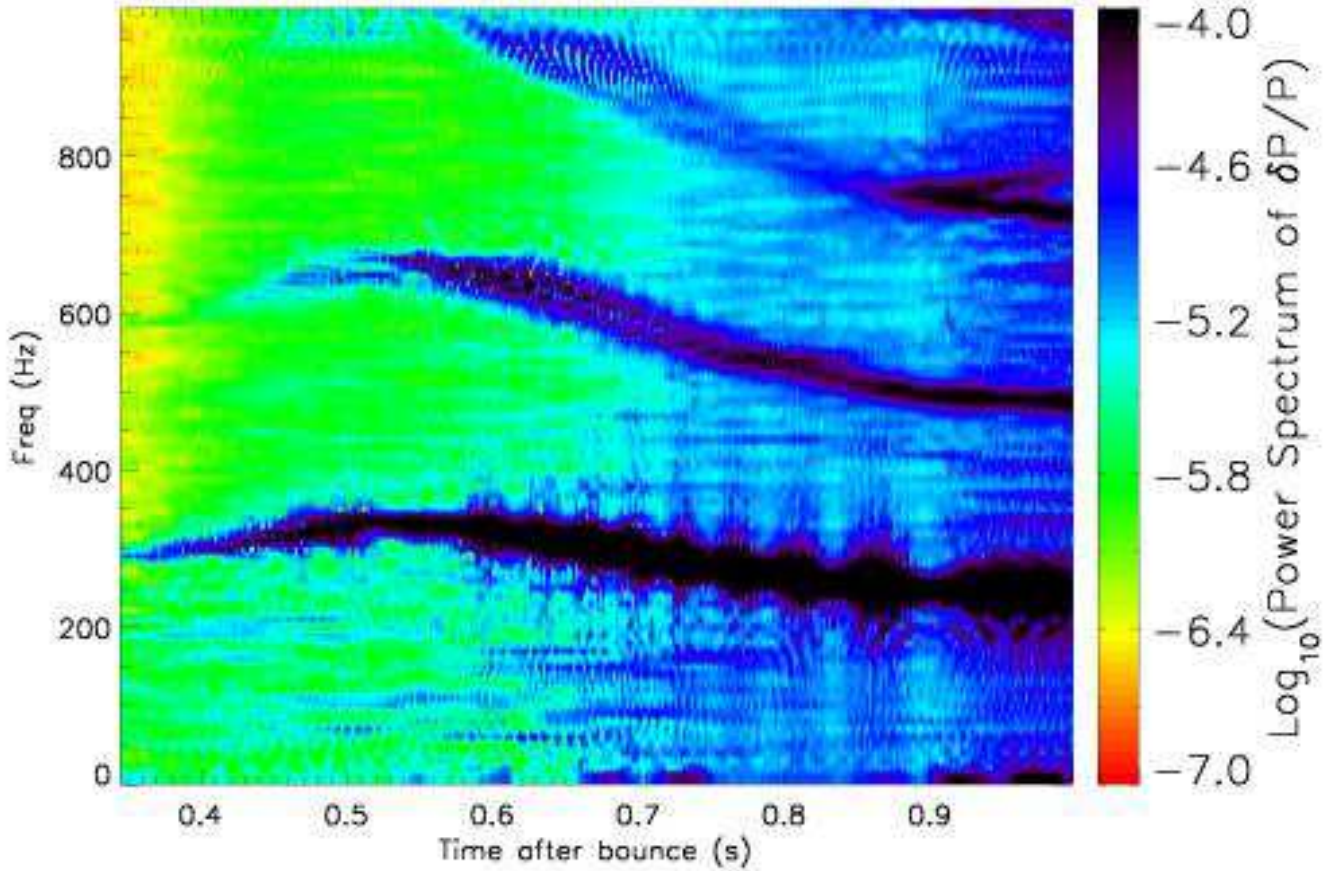, width=1.0\textwidth}
\caption{Colormap of the power spectrum of the fractional pressure variation
($P(R,\theta)-<P(R,\theta)>_\theta) / <P(R,\theta)>_\theta$
at a radius $R = 30$\,km, as a function of time after bounce and frequency.
For each time, a power spectrum is calculated from a sample of
time snapshots covering $t\pm$50\,ms, at a resolution of 0.5\,ms.
Note the emergence of power in the $\sim$330 Hz ($\equiv\,$3 ms) $\ell = 1$ g-mode,
as well as the strengthening at late times of 
the first harmonic oscillation near $\sim$675 Hz. The
$\ell=2$ component of the latter is of relevance for gravitational
radiation emission.  Note also the late-time evolution of the characteristic frequencies.
Figure taken from the study by Burrows et al. \cite{bur06}.}
\label{aburrows_freq}
\end{figure*}

\begin{figure*}
\epsfig{file=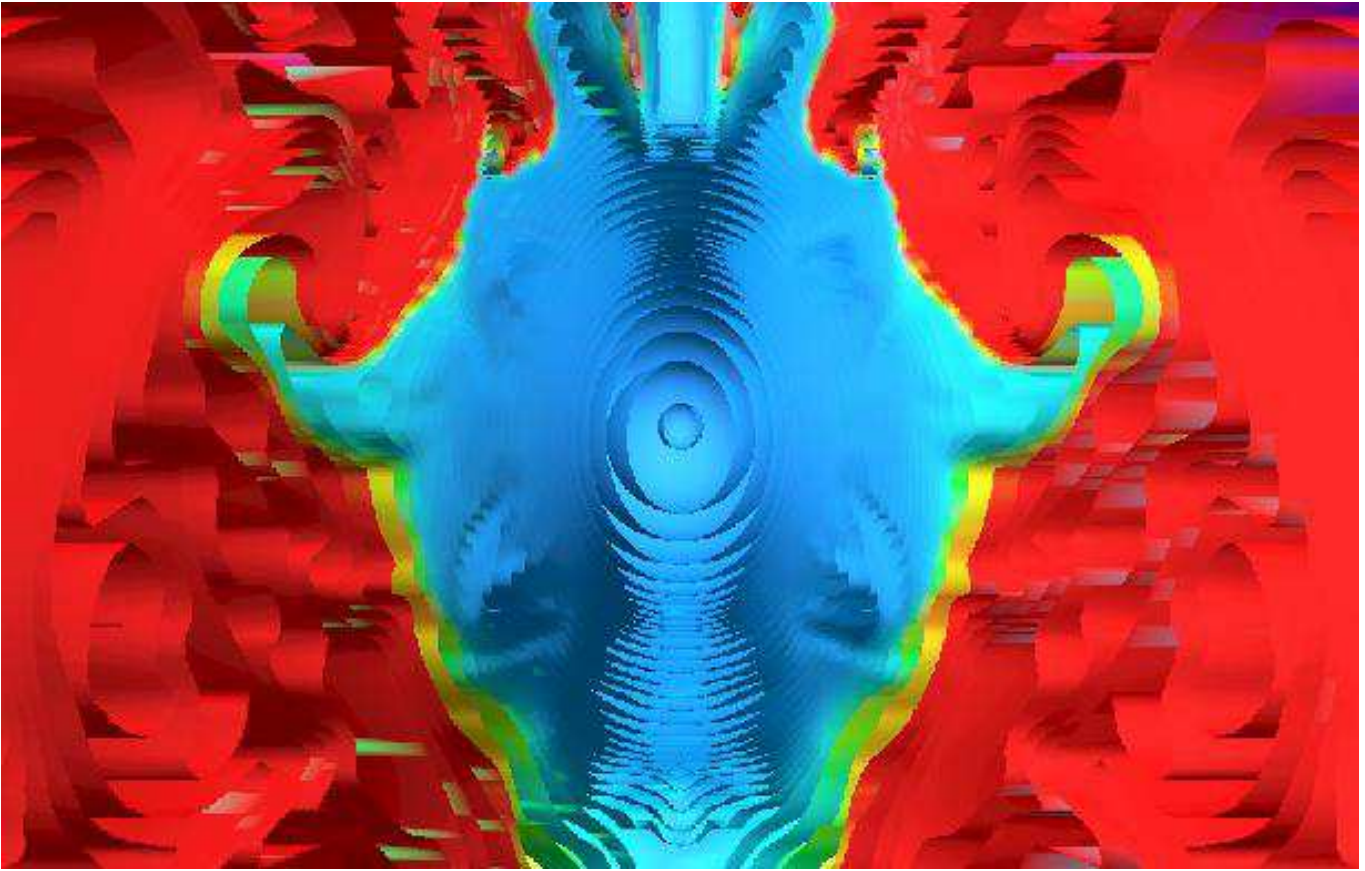, width=1.0\textwidth}
\caption{3D plot depicting isodensity sheets in the core of the protoneutron star (PNS)
at late times after the core oscillation has achieved significant amplitudes.
The linear scale is about 50 km.  The inner blue region is at low entropies and
the outer red region is turbulent and at much higher entropies.  The relative displacement
of the two inner shells due to the $\ell = 1$ g-mode is discernible.}
\label{aburrows_core}
\end{figure*}

\begin{figure*}
\epsfig{file=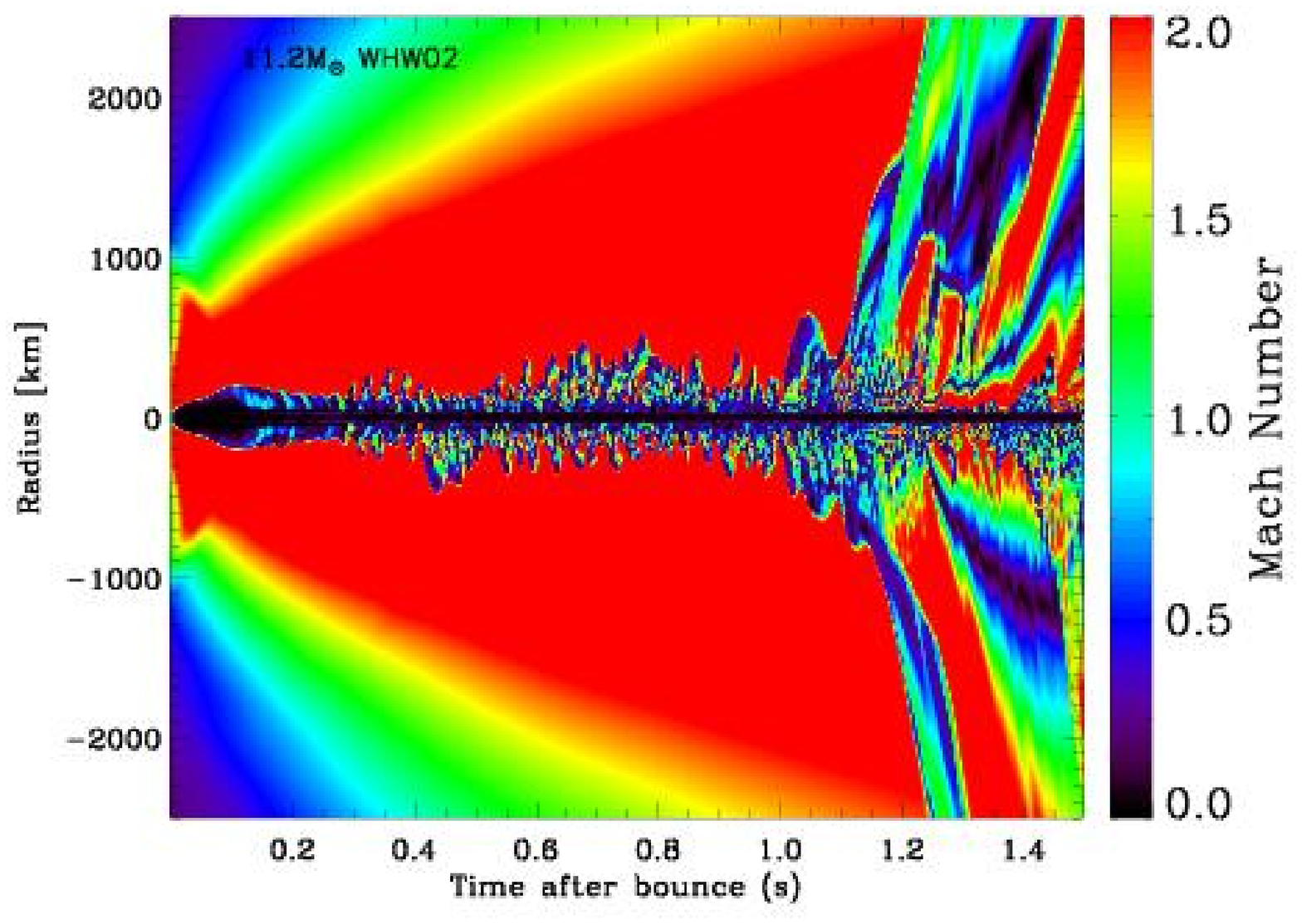, width=1.0\textwidth}
\caption{
Colormap of the time evolution after bounce of the Mach number profile along the poles (in both the positive
and the negative directions) for the 11.2-M$_{\odot}$
model of WHW02. The Mach number is saturated at a value of two to enhance the contrast.
The large red region is infalling material. The position of the bounce shock (near $\pm$200-500 km) 
and its oscillations before explosion due to the SASI are clearly seen, as are the multiple shocks that
emerge asymmetrically at late times to power the explosion.
}
\label{aburrows_mach}
\end{figure*}

\begin{figure*}
\epsfig{file=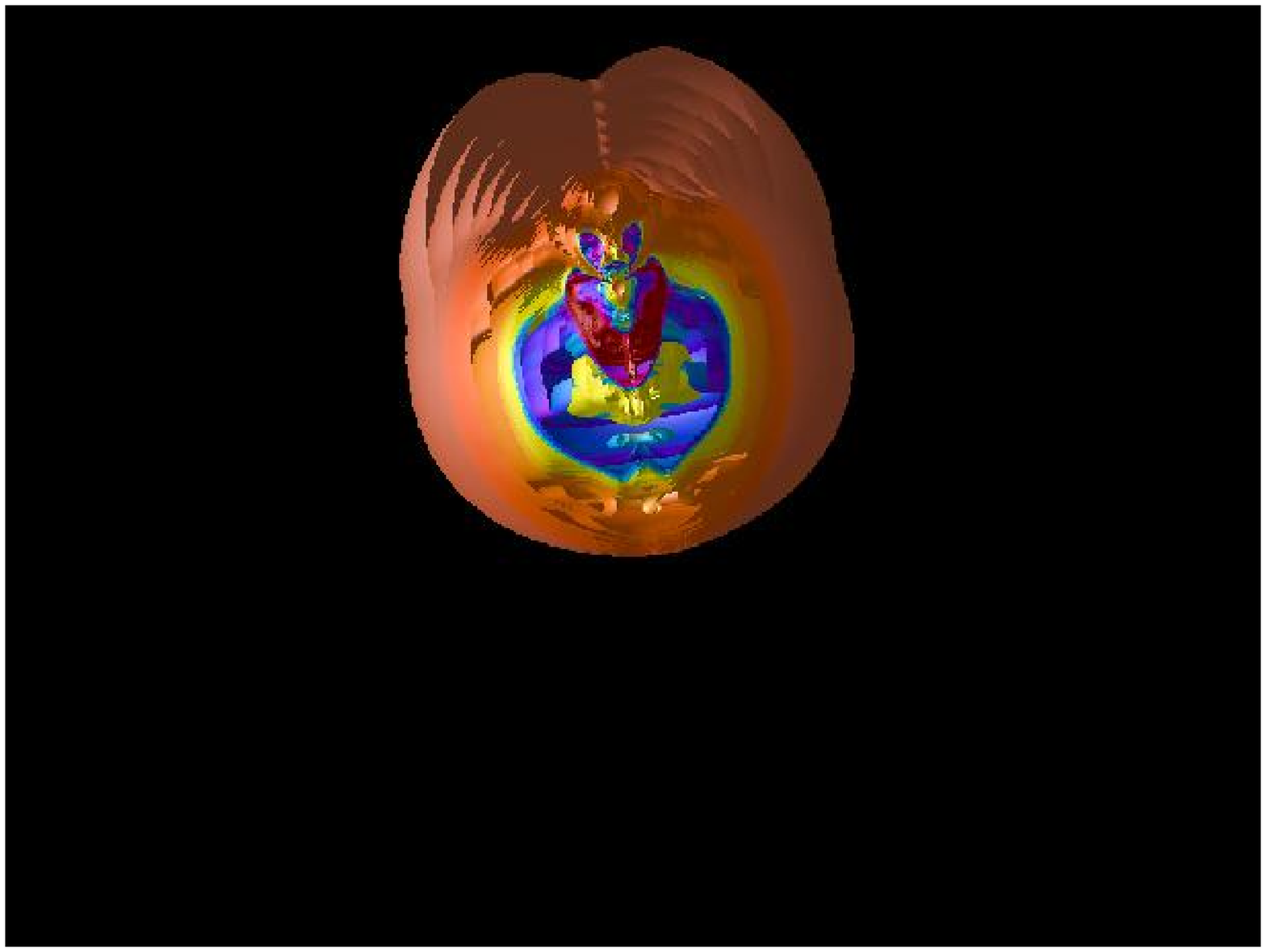, width=0.5\textwidth}  
\epsfig{file=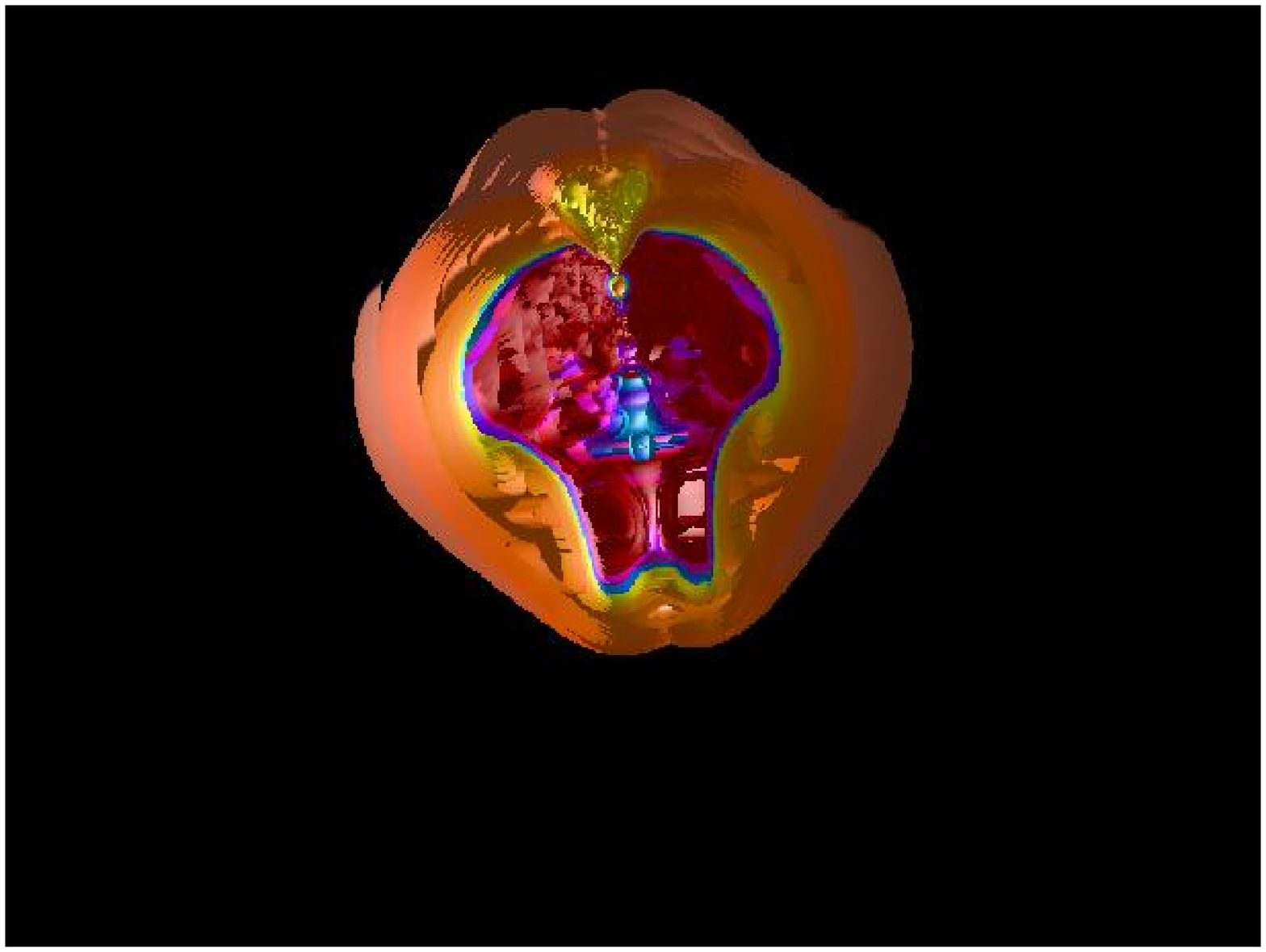, width=0.5\textwidth} 
\epsfig{file=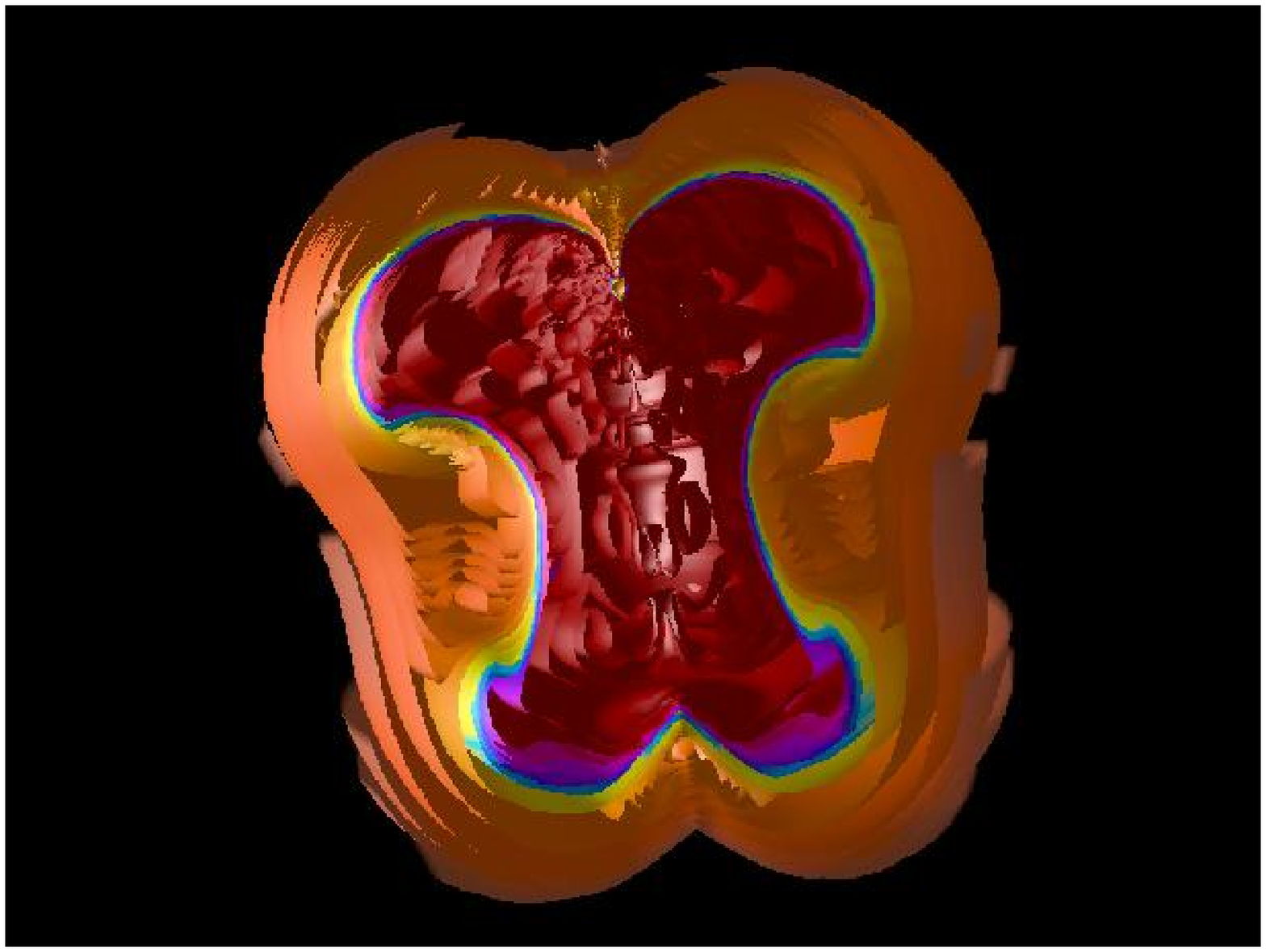, width=0.5\textwidth}
\epsfig{file=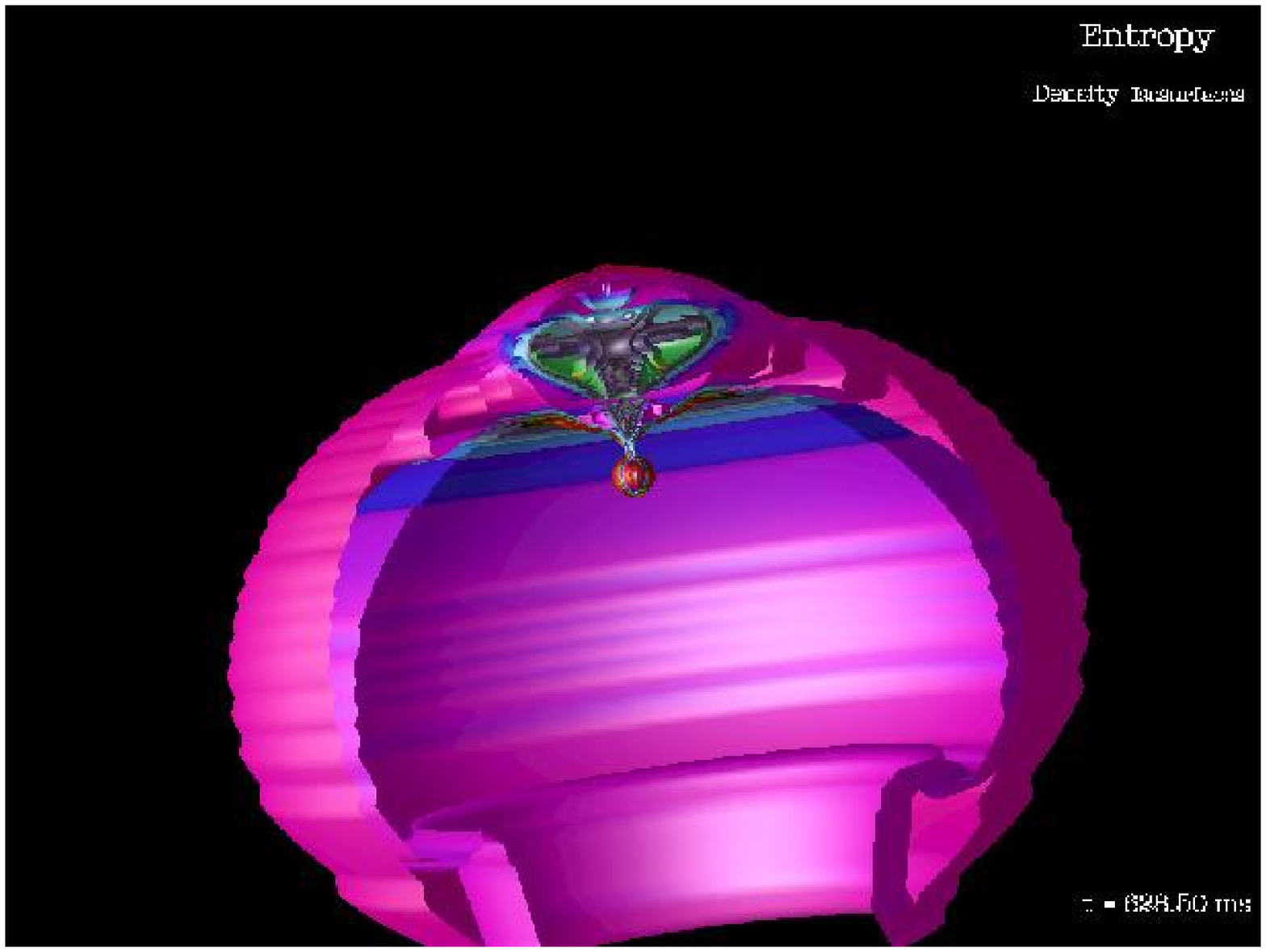, width=0.5\textwidth} 
\caption{
A temporal sequence of 3D renditions of isodensity sheets painted with entropy values
for a 2D model exploding by the anisotropic acoustic mechanism.  The explosion is predominantly
along one axis, during which accretion streams (here near the top) continue to impinge upon the core.
There is simultaneous accretion and explosion.
The central sphere is the protoneutron star and an expanding cavity is progressively excavated and filled with
acoustic power and thermal energy.  The scale for the first three plots is approximately 5000 km on a side,
while that for the last one is approximately 10000 km.
}
\label{aburrows_nuts}
\end{figure*}

If the neutrino mechanism is eventually shown to work in the 
first few hundred milliseconds after bounce, perhaps due to 3D 
effects or better physics and numerics, this will be welcome news 
to supernova theorists. However, what if most of the time it does not? 
Recently, Burrows et al. \cite{bur06,bur06b}, using the code VULCAN/2D, have suggested an explosion driver  
that relies predominantly on acoustic power generated by vigorous core pulsations.
First, within $\sim$200 ms after bounce the stalled shock experiences 
the SASI with a period of $\sim$15$-$30 milliseconds (ms) and begins to 
execute large deviations from sphericity.  The growth of this 
outer shock instability saturates due to the generation of secondary 
shocks.  Our calculations support the notion that all 
non-rotating progenitors that do not explode by an early neutrino 
mechanism experience the SASI, whose oscillation periods are best interpreted simply
as the sound-travel-times across the shocked regions.  However, this instability is not the primary 
agent of explosion.  Rather, it is the acoustic power generated early on 
in the inner turbulent region stirred by the SASI-generated accretion plumes,
and most importantly, but later on, by the consequent excitation and
sonic damping of rapid core g-mode pulsations.  An $\ell=1$ g-mode
(not the SASI) with a period of $\sim$3 ms grows at late times to
be prominent around $\sim$500 ms after bounce.  Figure \ref{aburrows_freq} portrays
the temporal evolution of core oscillation frequencies for a representative model calculation. 
%
Figure \ref{aburrows_core} provides a 3D snapshot in isodensity sheets of the $\ell = 1$ dipolar
pulsation of the inner core in a calculation from Burrows et al. \cite{bur06,bur06b}.  
The accreting protoneutron star (PNS) then acts like a self-excited oscillator (see \S\ref{excite}).
The sound pulses radiated from the core steepen into shock waves that
merge as they propagate into the outer mantle and deposit their energy
and momentum with high efficiency.  The ultimate source of the acoustic
power is the gravitational energy of infall. We find that the associated acoustic power is sufficient to drive                                
the explosion of all models studied to date $>$500$-$1100 milliseconds after bounce.
The delay to explosion and its sudden onset is depicted in Fig. \ref{aburrows_mach}.  The colors
represent the Mach number of the matter.  Note that the Mach number of the turbulence behind the shock
increase with time, and that the secondary shock waves that emerge during the non-linear phase of the
SASI and during the explosion are clearly in evidence.
%
The angular distribution of the emitted sound is fundamentally aspherical
and the acoustic power fills a growing cavity that drives the explosion.
The evolution of such a cavity for a representative core-acoustic 
explosion is presented in Fig. \ref{aburrows_nuts}.
%
An advantage of the acoustic mechanism is that acoustic power does
not abate until accretion subsides, so that it is available as long
as it may be needed to explode the star.

There are certain other virtues to acoustic driving that bear mentioning.
First is that while the acoustic luminosity is much smaller than the neutrino
luminosity, almost all of the sound is absorbed in the mantle matter. At late
times in our simulation, less than a percent of the
$\nu_e$ and $\bar{\nu}_e$ neutrino luminosity is absorbed.
This amounts to an neutrino absorption power of $\le$$10^{50}$ erg s$^{-1}$,
compared with an estimated core acoustic power at the end of our calculations
near $\sim$${10}^{51}$ erg s$^{-1}$. Second, sound
carries not only energy, but momentum, and this factor seems to be important in our
simulations.  The momentum flux for sound with the same energy flux as neutrino radiation
is larger by the ratio of the speed of light to the speed of sound, which in the
inner mantle regions is as much as a factor of ten.   Third, acoustic power propagates
from where it is generated to where it is needed; it fulfills the central requirement
of a core-collapse supernova mechanism that it involve energy transfer from the bound
interior PNS to the outer exploding mantle.  If the acoustic power is large
enough, it is the ideal transfer agent.  Fourth, the acoustic source
seems to grow just when the neutrino luminosity is ebbing and, importantly, it
continues until explosion ensues.  Fifth, the successive merger of trains of sound waves
that steepen into shocks provides a non-neutrino way to entropize some of the matter
and naturally achieve r-process conditions.

It should be pointed out that the SASI and the breaking of spherical symmetry are likely  
important even in the context of the neutrino-driven explosion scenario, and in fact
in every viable explosion scenario that is emerging from modern calculations.
When the SASI is in its vigorous non-linear phase, its $\ell = 1$ oscillations
result in quasi-periodic fluctuations in the effective accretion rate and ram pressure on any given
side of the inner core.  In the canonical neutrino-driven mechanism of supernova
explosions, when and after the explosion occurs the pressure around the neutrinospheres
decays.  When this pressure is sufficiently low, a neutrino-driven wind spontaneously
emerges from the inner core, announced and preceded by a secondary shock wave \cite{bg93,bhf}. 
This is what happens in the standard neutrino-driven scenario when the flow is semi-spherical. However, the SASI 
can set up a situation in which the pressure and ram pressure on one side 
render that side of the core unstable to the emergence of 
a neutrino-driven wind even before the canonical explosion. In fact, this wind can
be the explosion itself and need not be preceded  
by a primary explosion.  This is what Buras et al. \cite{buras2} see for their 11.2-M$_{\odot}$  
simulation.  However, such an explosion seems generically underenergetic. 
In the acoustic mechanism, the neutrinos are replaced/dominated by the acoustic power,
but the general paradigm in which the SASI leads temporarily/periodically to  
lower pressures on one side of the core that enable the emergence of an asymmetric wind
still obtains.  In any case, an aspherical ``wind" is a good description 
of the supernova explosion \cite{bur06b,scheck,scheck2,bg93}  
and {\bf the breaking of spherical symmetry is the key}.  The 
latter can also enable simultaneous accretion and explosion,
thereby solving the problem of the accretion tamp that has bedeviled  
the theory of the neutrino mechanism for years.

\begin{figure*}
\epsfig{file=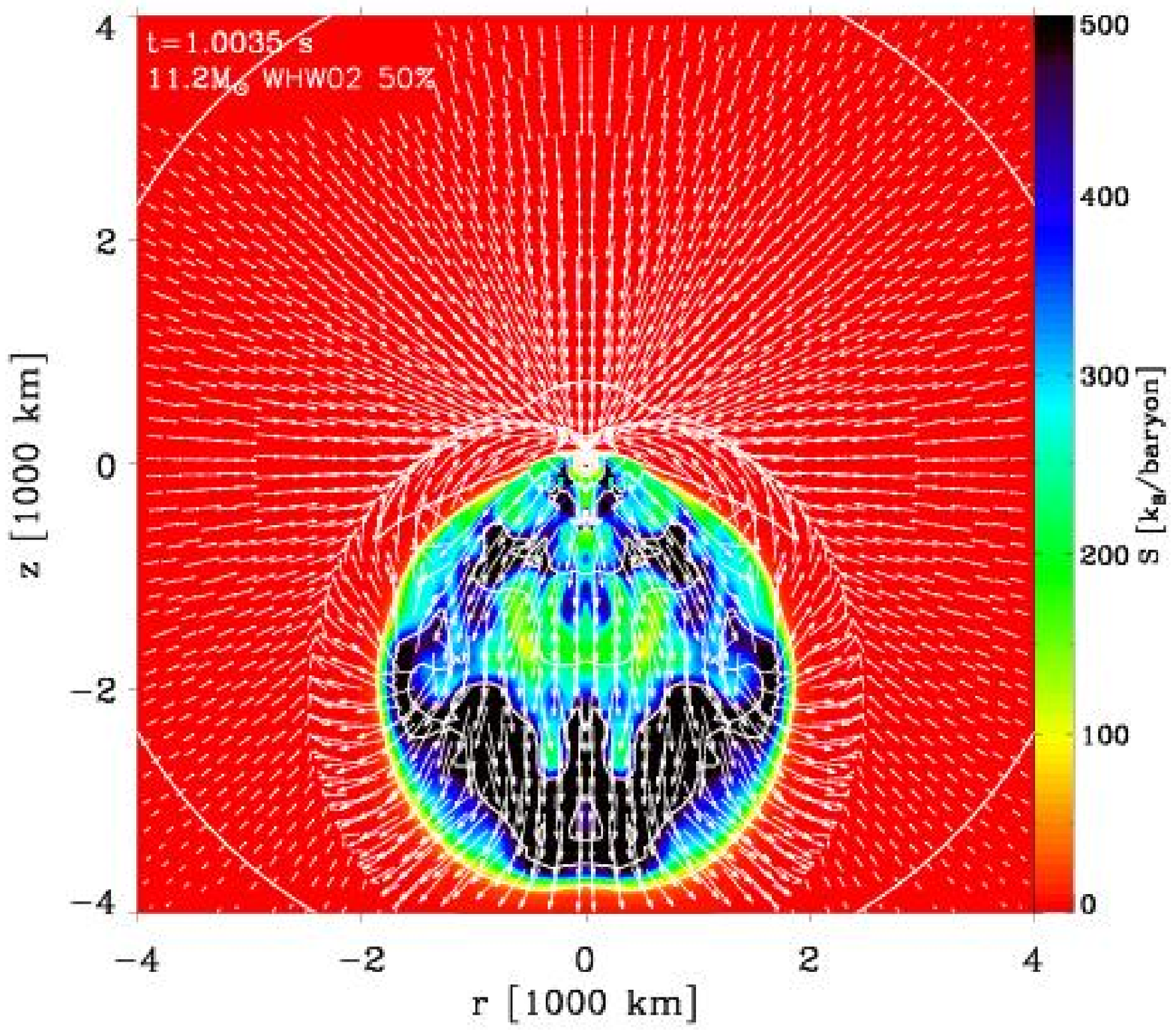, width=0.5\textwidth}  
\epsfig{file=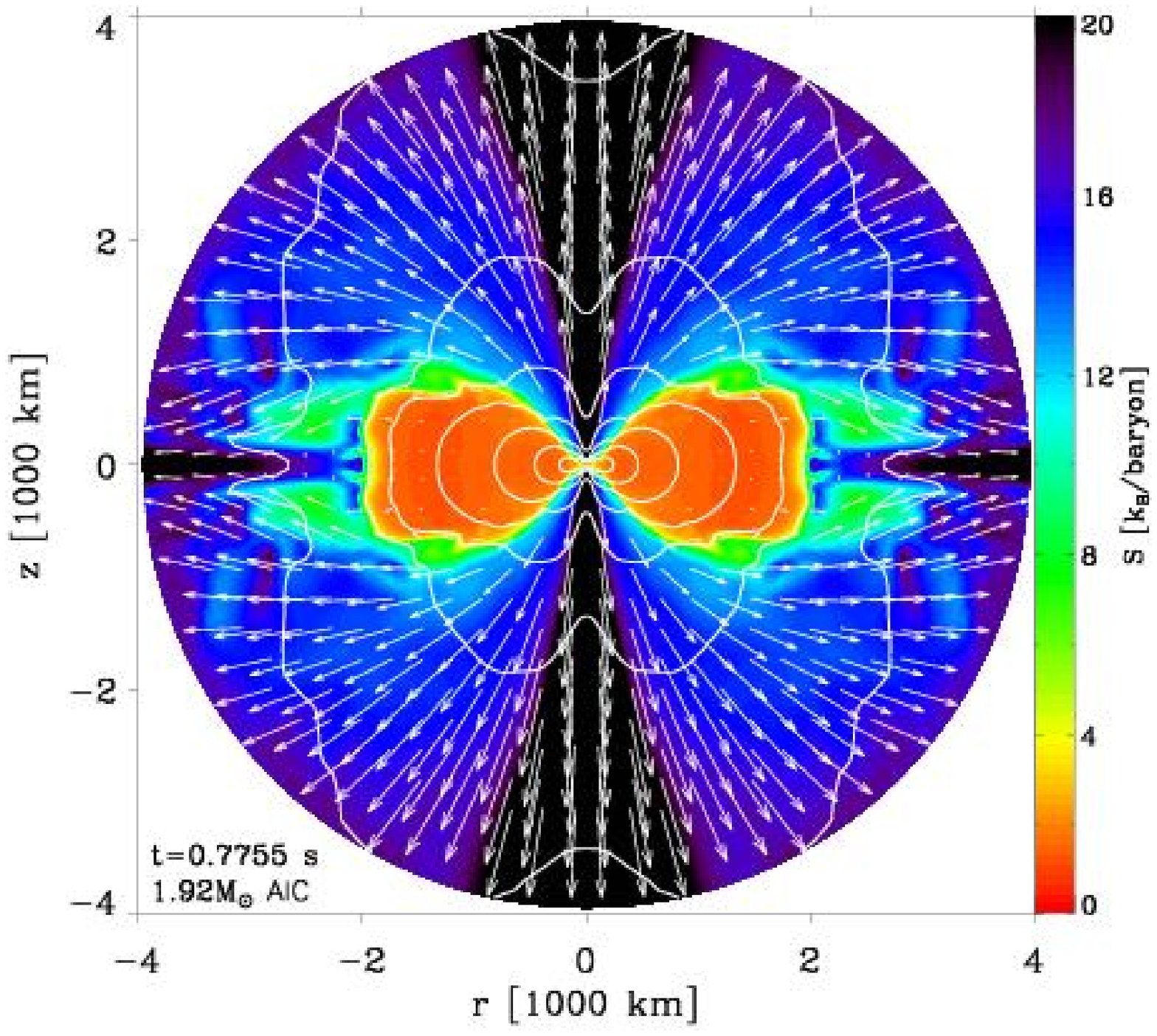, width=0.5\textwidth}  
\epsfig{file=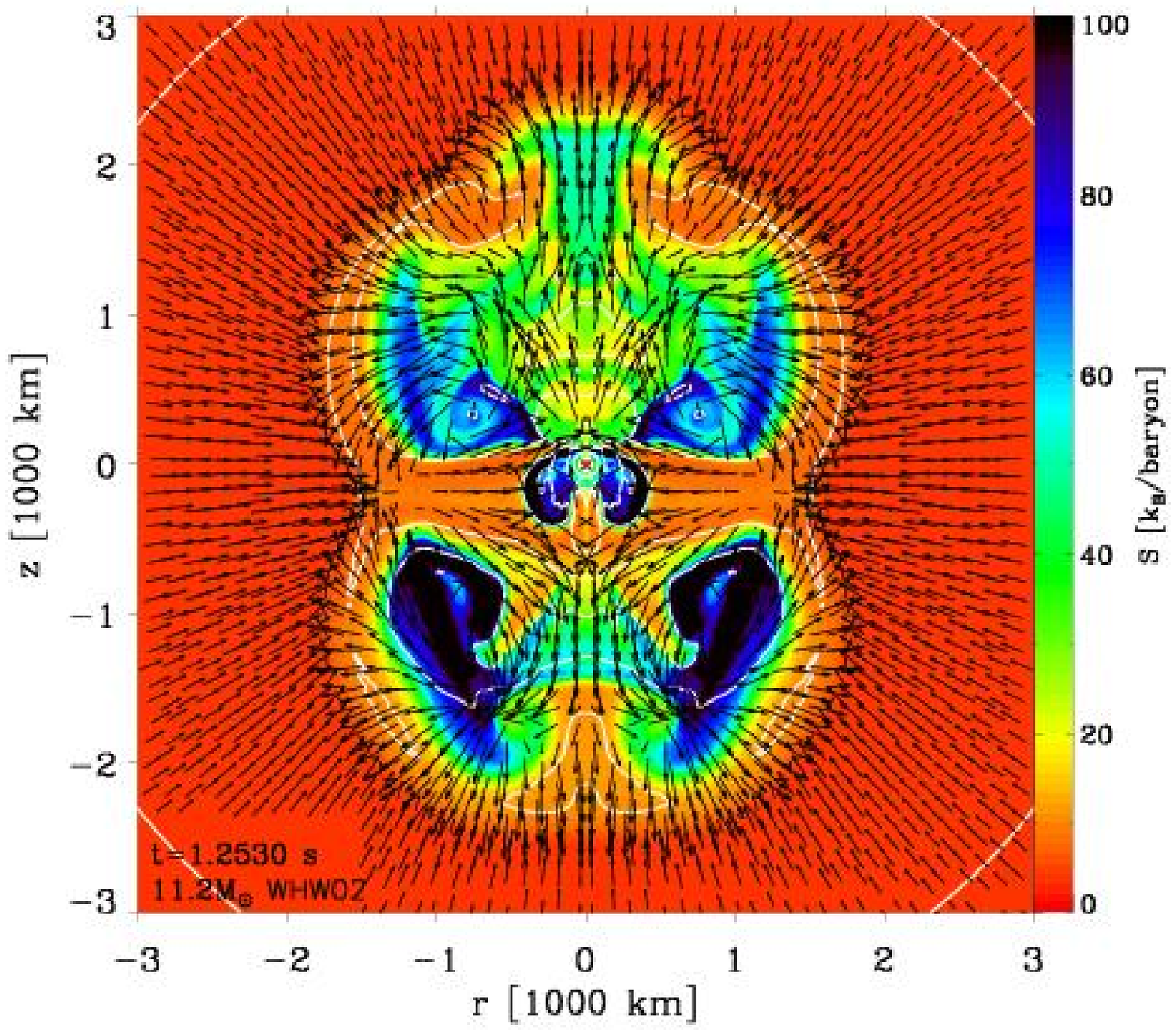, width=0.5\textwidth}  
\epsfig{file=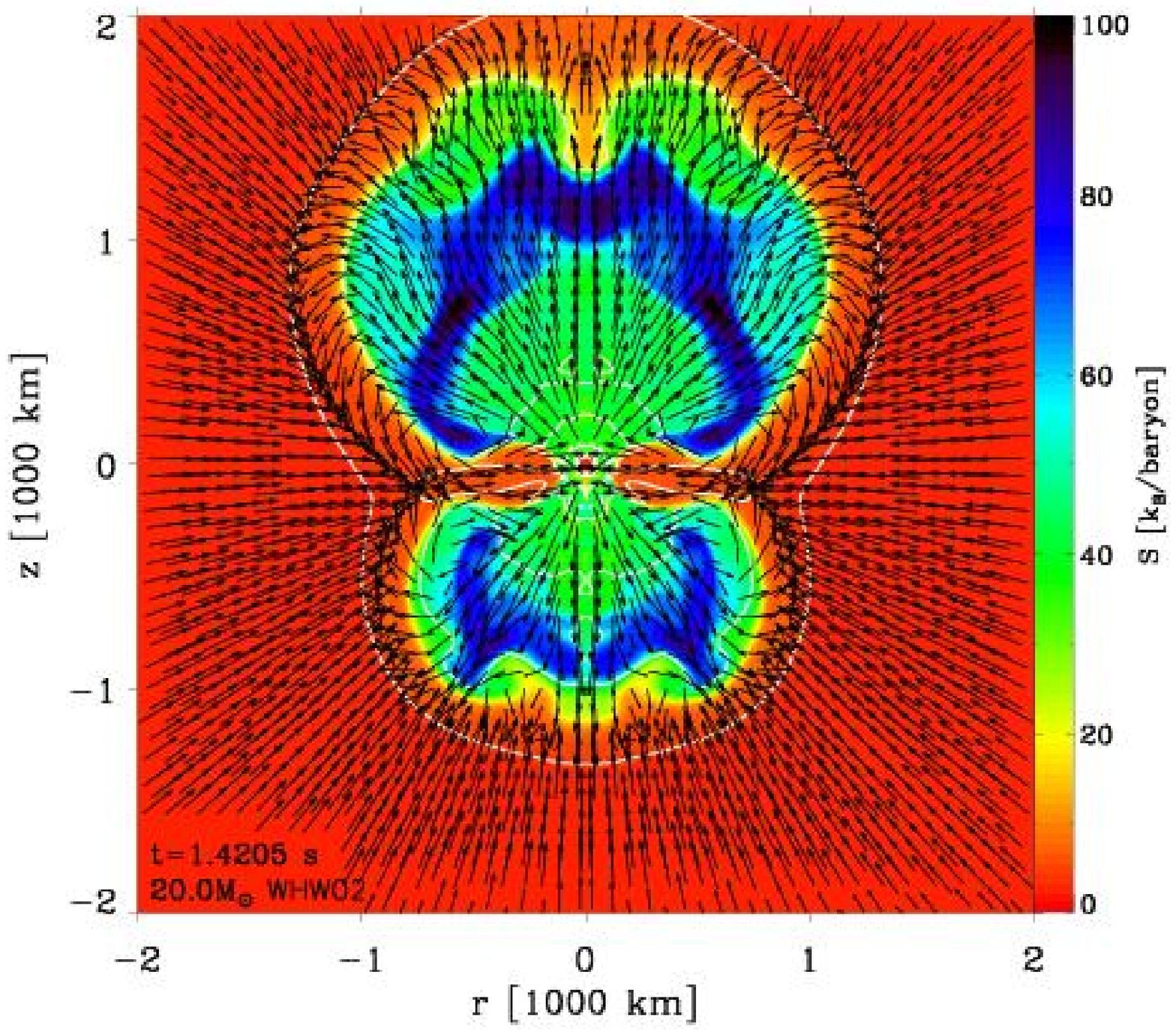, width=0.5\textwidth}  
\caption{
Colormap of the entropy for our simulation of the 11.2-M$_{\odot}$ model of WHW02, but
with the neutral-current scattering rates diminished artificially by 50\%
(top left), the 1.92-M$_{\odot}$ AIC model of
Dessart et al. \cite{dessarta} (top right), the default 11.2-M$_{\odot}$ model of WHW02
(bottom left), and the 20-M$_{\odot}$ model of WHW02 (bottom right).
Times after bounce are indicated in each panel.
The maximum length vector, set to 5\% of the width of the display, corresponds
to a velocity of 10000 km s$^{-1}$ (values above that are saturated), except for the AIC model for which
the corresponding values are 10\% and 20000 km s$^{-1}$.
}
\label{aburrows_morph}
\end{figure*}

Hence, we see in the breaking of spherical symmetry in our simulations
and in the often unipolar nature of the resulting explosions a natural explanation for the
polarizations observed in the inner debris of Type Ic \cite{wang2} and Type II \cite{leonard} supernovae.
Inner asymmetries of 2:1 or 3:1 are easily obtained in this model and do not
require MHD jets.  Figure \ref{aburrows_morph} compares explosion morphologies
for a representative range of VULCAN/2D simulations.  The actual degree and
character of the anisotropy is unpredictable, given the chaotic dynamics, but spherical
explosions should be exceedingly rare.
%

\subsection{Excitation Spectrum of Gravity Waves during the Late-time Anisotropic Accretion Phase}
\label{excite}

The explosions we see resemble strong anisotropic winds.
A spherical wind imparts no net momentum to the residue; an asymmetric wind
imparts a kick and ``ablation" force on the accretion streams and core.
The recoil implied is a purely hydrodynamical mechanism, whatever the
agency of explosion (be it neutrinos or sound), and has two results.
First, the recoil due to the anisotropic wind pushes the accretion streams
to the opposite side, making the accretion very anisotropic.  A fraction
of the gravitational energy of accretion is used to continue to excite the inner
core g-mode oscillation.  Because the accretion funnels are supersonic,
the coupling to the core is non-linear.  Importantly, the oscillation
of the core can not do work back on the exciting accretion stream(s) that
would otherwise damp the core oscillation; any work done is accreted back. Hence, the analogy
with the swing which requires a resonance or near resonance to achieve
significant amplitude is not germane.  A steady stream onto the core
can continue to power the periodic core oscillation, even though there
is no intrinsic periodicity to the accretion.  The accretion funnel
does have a width, which like a rock hitting a pond has associated with
it a range of characteristic sizes (read wavelengths).  Due to the
dispersion relation of gravity waves between wavelength and frequency, a whole period spectrum
of ripples is generated which contains the period of the $\ell = 1$ core g-mode
(as well as those of many of the higher-$\ell$ core g-modes).  

For instance, if the steady accretion stream has a width of $\sim$10 km, the average frequency 
of the gravity waves generated at the collision with the inner core whose surface
gravity is $\sim$10$^{13}$ cm s$^{-2}$ is $\sim$700 Hz, with a spectrum that extends
from lower to higher frequencies that comfortably overlaps with the core g-mode eigenfrequencies.
This is true despite the fact that there is no such periodicity in the exciting 
accretion stream itself.  The thinner the stream the wider the excitation frequency spectrum.
This limit is analogous to the ``white noise" spectrum of gravity waves excited by the narrow bow
of a ship that collectively result in the characteristic scale-free, fixed-angle bow wave.
What is not clear from merely analytic arguments is the fraction of the accretion
power that should be channeled into mechanical energy.  The fraction channeled into neutrino losses
is usually the larger. One can estimate in the crudest fashion the power being pumped
into gravity waves as a function of the stream radius, $R_0$, the wave height, $H_0$, the average
density at the core ``surface," $\rho$, and the surface gravity, $g$:

\begin{equation}
\dot{E} \sim \frac{\pi\rho}{2} (gR_0)^{3/2} H_0^2 \sim 0.5\times 10^{51} {\rm erg\, s}^{-1} \rho_{11}\left(g_{13}\right)^{3/2} \left(\frac{R_0}{10 {\rm km}}\right)^{3/2} \left(\frac{H_0}{3 {\rm km}}\right)^2\, ,
\label{eqedot}
\end{equation} 

where $\rho_{11} = \rho/(10^{11}\, {\rm g\, cm}^{-3})$ and  $g_{13} = g/(10^{13}\, {\rm cm\, s}^{-2})$.  
The reference value of $\dot{E}$ in eq. (\ref{eqedot}) is suggestive. To estimate $H_0$ 
we would need the fraction of the accretion power that should be set equal to $\dot{E}$, something
currently we can get only from imperfect numerical simulations.  However, the reference 
numbers in eq. (\ref{eqedot}) come from such simulations.

\section{Angular Distribution of Neutrino Flux and Possible Kick Mechanisms}
\label{aniso_flux}

With VULCAN/2D, Burrows et al. \cite{bur06b} are able to ascertain the
magnitude and sign of the impulse due to anisotropic neutrino emissions. We find that
during our simulations (to approximately 1.5 seconds after bounce)
the neutrino recoil effect on the core is not large, at most $\sim$50 km s$^{-1}$, but that by
the end of our simulations it is still growing and is in the opposite direction
to the blast.  Moreover, after the explosion commences, the impulses on the
protoneutron star due to the matter ejecta and the neutrino radiation {\it add}.
The small magnitude of the neutrino force during the delay to
explosion may seem inconsistent with the very anisotropic accretion.
%
\begin{figure*}
\epsfig{file=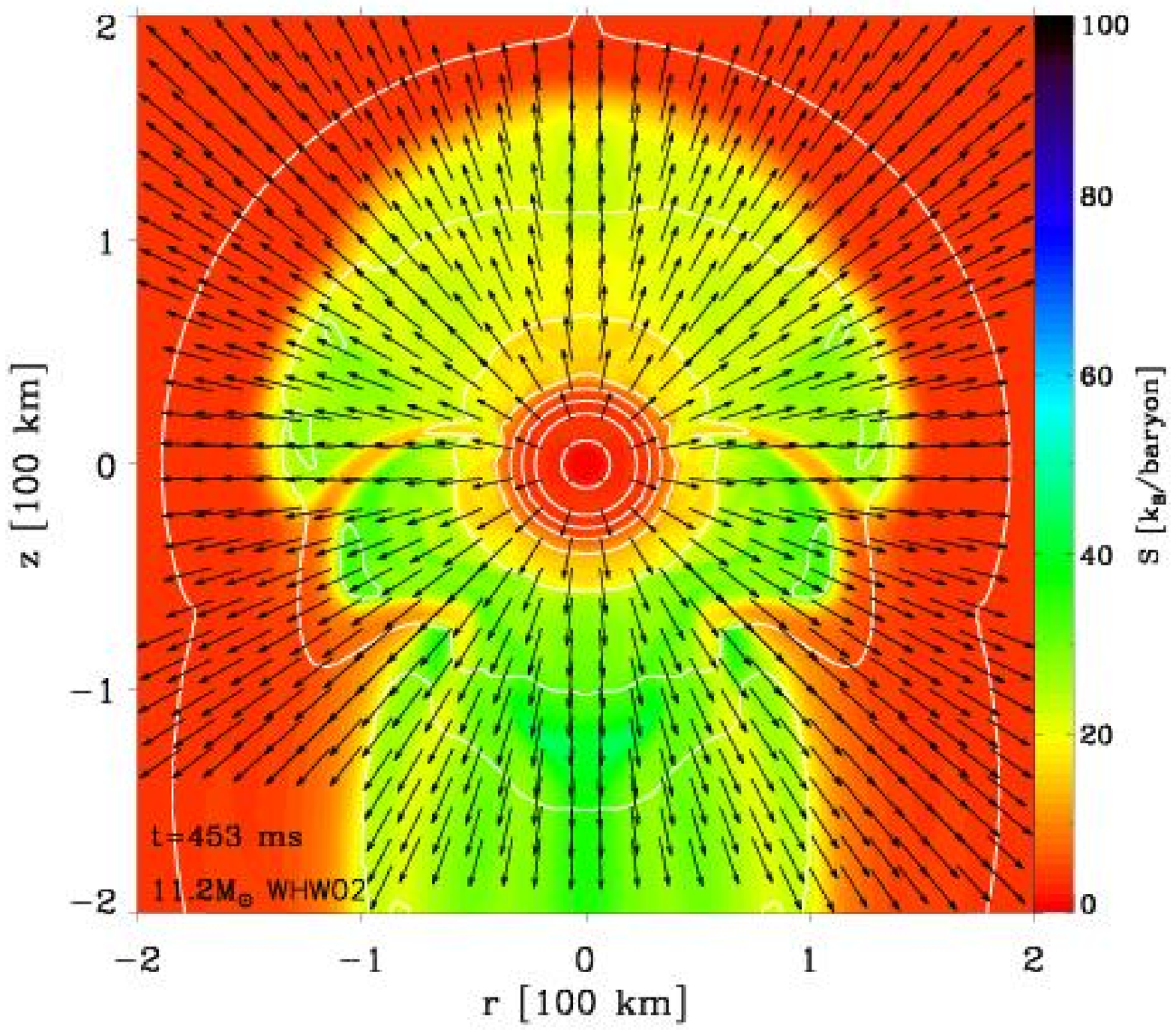, width=0.5\textwidth}  
\epsfig{file=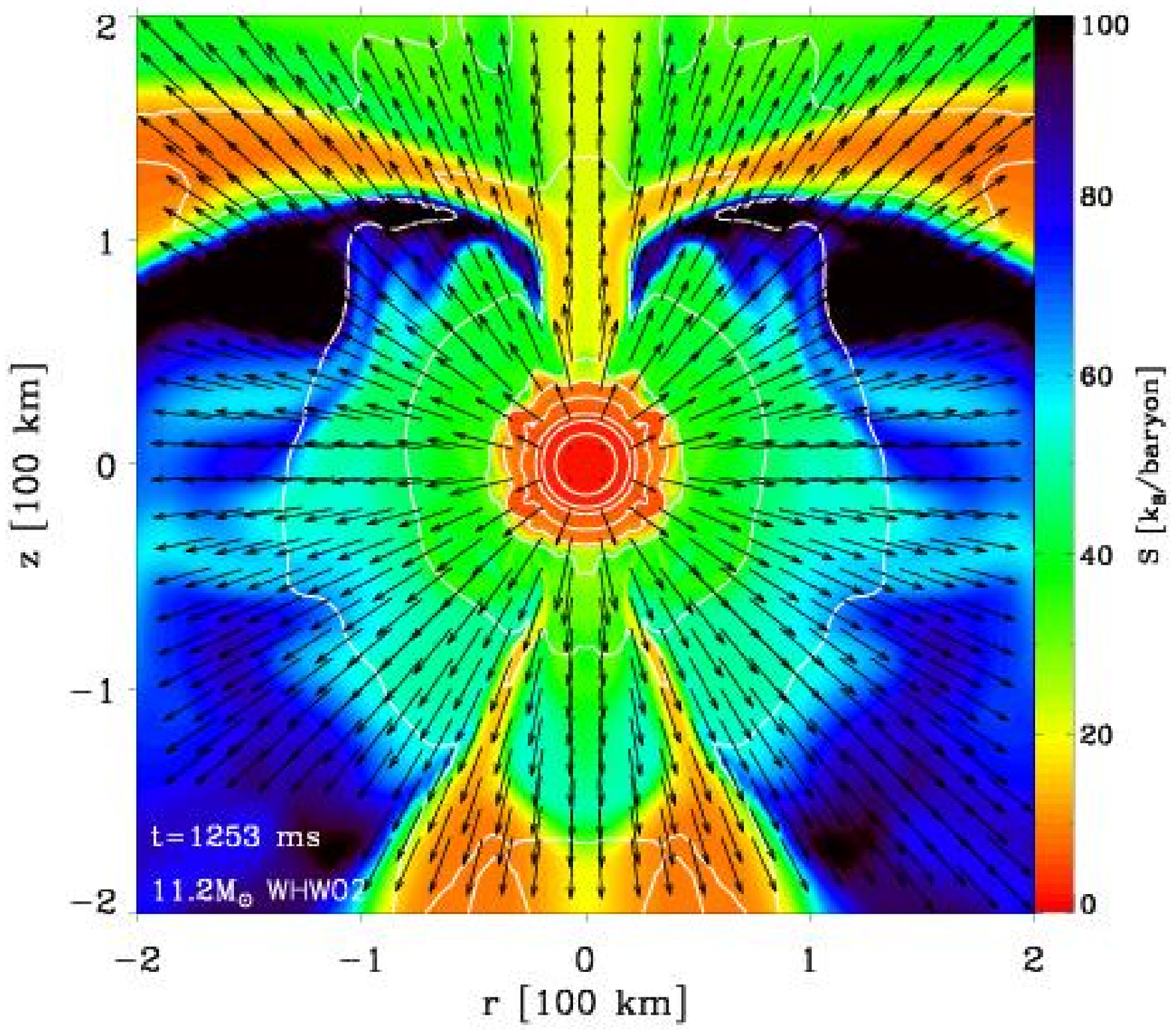, width=0.5\textwidth}  
\epsfig{file=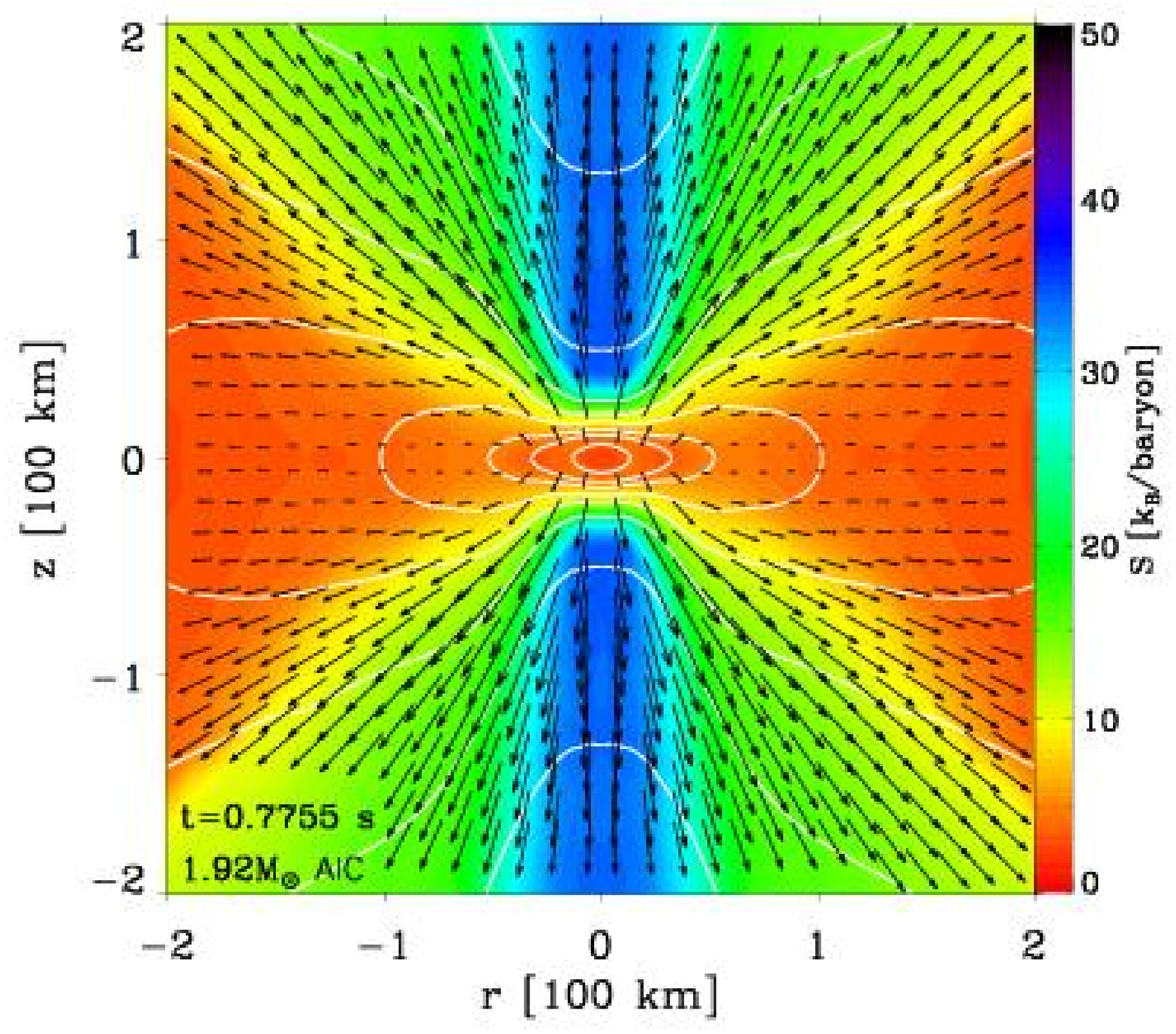, width=0.5\textwidth}  
\caption{
Colormap of the entropy for the 11.2-M$_{\odot}$ model of WHW02,
at 453\,ms (left) and 1253\,ms (center) after bounce \cite{bur06b}, and for the
1.92-M$_{\odot}$ Accretion-Induced Collapse (AIC) white
dwarf model of Dessart et al. \cite{dessarta} at 775.5\,ms after bounce (bottom).
Density contours are also drawn in white, for every decade down from
10$^{14}$\,g\,cm$^{-3}$ (innermost and spherical ring).
We overplot vectors of the quantity $4 \pi R^2$ $\vec{F}_{\nu}$
(where $\vec{F}_{\nu}$ is the total neutrino flux), and saturate
this quantity at $1.2\times 10^{53}$erg\,s$^{-1}$ with a maximum length of 5\% of the display width.
Note how the neutrino flux remains essentially radial in the non-rotating
model (left and center), despite the large deviations
from spherical symmetry in the entropy and density distributions.
By contrast, in the fast rotating model of the AIC of Dessart et al. \cite{dessarta} significant
asphericity prevails in the high-density regions which modify considerably the
emergent neutrino flux, enhanced along the pole, but diminished along the equator.
}
\label{aburrows_flux}
\end{figure*}
However, the radiation field is much smoother by its nature than the material
field. Figure \ref{aburrows_flux} provides vector fields of instantaneous 
neutrino flux distributions superposed on background entropy maps for three different models.
Only for the accretion-induced-collapse model (see \S\ref{aic}) is the flux very aspherical,
and this is due to the very rapid rotation. For the non-rotating models, the 
neutrinos are not radiated instantaneously upon compression in an
accretion column onto the protoneutron star.  The matter is too opaque for immediate
reradiation.  Rather, the neutrinos emerge after the compressed accreta have
spread more uniformly over the inner core, and, therefore, are radiated much more isotropically
than the matter is accreted.  But importantly, after explosion the neutrinos can emerge more easily
along the direction of the blast, since the material around the neutrinospheres
thins out in this direction, and not along the direction experiencing continuing accretion,
which, as stated, is more opaque.  
%
\begin{figure*}
\epsfig{file=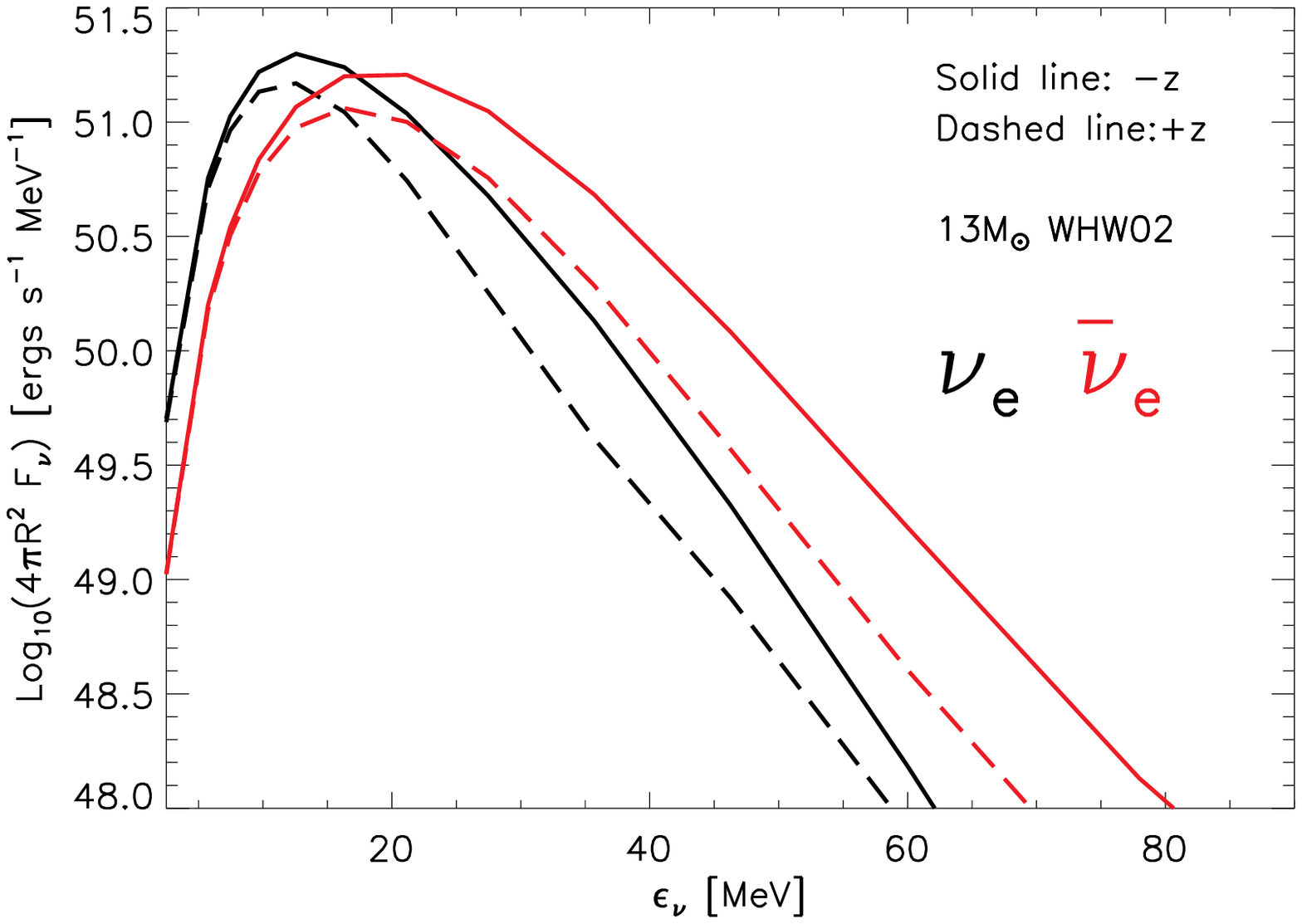, width = 0.5\textwidth}
\epsfig{file=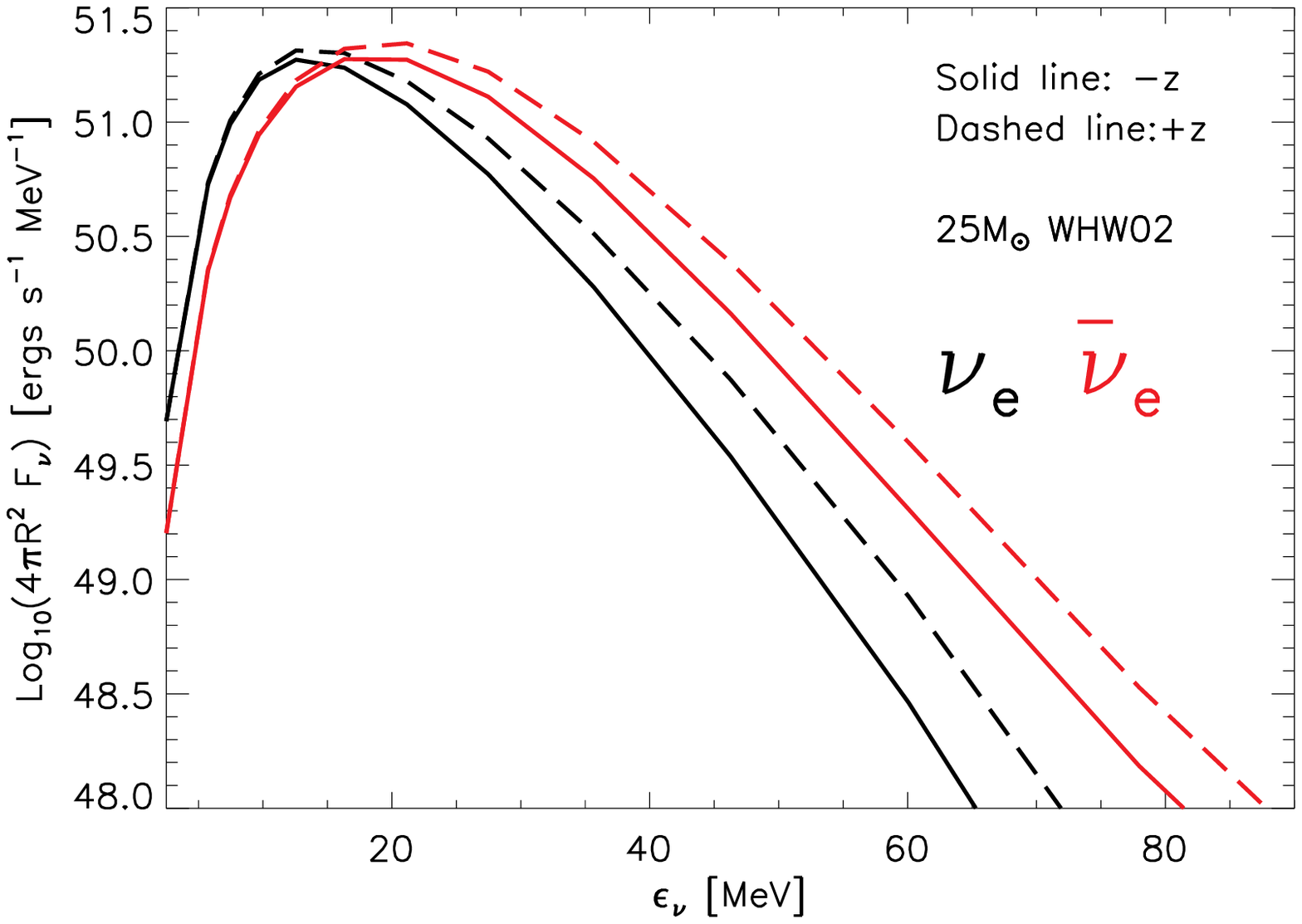, width = 0.5\textwidth}
\caption{{\it Left:} Energy spectra of the electron-neutrino (black) and
anti-electron neutrino (red) fluxes along the poles (solid line: negative
$z$-direction; dashed line: positive $z$-direction) for the 13-M$_{\odot}$
model of WHW02.  The fluxes are multiplied by a factor $4 \pi R^2$ and
are at 1.44\,s after bounce and after the explosion commences. {\it Right:} Same as at the left, but for
the 25-M$_{\odot}$ model of WHW02 at 1.42\,s after bounce. Note that the hotter and
higher fluxes are in each case in the direction of the explosion, though the two
models shown explode in different directions.
Figure taken from Burrows et al. 2006b.}
\label{aburrows_spectra}
\end{figure*}
Figure \ref{aburrows_spectra} depicts the spectra of the
$\nu_e$ and $\bar{\nu}_e$ neutrinos in the up and down
directions (along the poles) near the end of the simulations of
the 13-M$_{\odot}$ and 25-M$_{\odot}$ models of WHW02.  As Fig. \ref{aburrows_spectra} demonstrates,
the radiation is ``hotter" in the direction of the blasts.  It is also ``brighter"
in those directions.  Hence, we suggest that asymmetric neutrino recoil, integrated 
over a long time after explosion, can contribute significantly and naturally
to the final pulsar kick.  

In fact, wind/explosion recoil and the neutrino emission anisotropy after explosion commences
provide natural mechanisms for imparting kicks to the residual neutron stars. The anisotropic/top-bottom
explosion acts like rocket exhaust and momentum conservation does the rest.
The magnitude of the effect can be approximated as follows: the recoil
force is equal to $\Phi{v}\dot{M_e}$, where $\Phi$ is the
average ``anisotropy parameter," $v$ is the characteristic wind velocity,
and $\dot{M_e}$ is the wind mass loss rate. The power poured
into the supernova ``nebula" by the wind is $1/2 \dot{M_e}v^2$.  Integrating
both these quantities over time gives the net impulse and explosion energy ($\eta$),
respectively.  The impulse is equal to the residue mass ($M_{pn}$) times the kick velocity ($v_k$).
Taking the ratio of these two expressions results in a formula for the kick velocity:
\begin{equation}
v_k = 2\eta /(M_{pn}v)\Phi \, .
\end{equation}
If we assume that the scale of $v$ is set by a sound speed ($\sim$30,000-100,000 km s$^{-1}$), we derive that
$v_k \sim 1000 (\eta /10^{51} {\rm ergs}) \Phi$ km s$^{-1}$. The average
observed/inferred kick speed is 300-400 km s$^{-1}$ \cite{cordes,lyne}, so this number is tantalizing.
The anisotropy parameter, $\Phi$, can be large, but depends on the stochasticity of the flow.
This formula works whether the explosion is driven by neutrinos or sound, and depends
only on the wind-like character of the asymmetric explosion and simple momentum conservation.

However, we can not, at this stage, determine
whether the matter recoil or the neutrino recoil will eventually prove the more important.
Nevertheless, what has emerged from our simulations is a straightforward mechanism,
matter {\it plus} neutrino recoils, for imparting a sizable kick to the residue protoneutron star that does not
require anything but the breaking of spherical symmetry and the asymmetric explosion that arises naturally
without exotic physics.

\eject

\section{Accretion-Induced Collapse: Rapidly-Rotating Models}
\label{aic}

\begin{figure*}
\epsfig{file=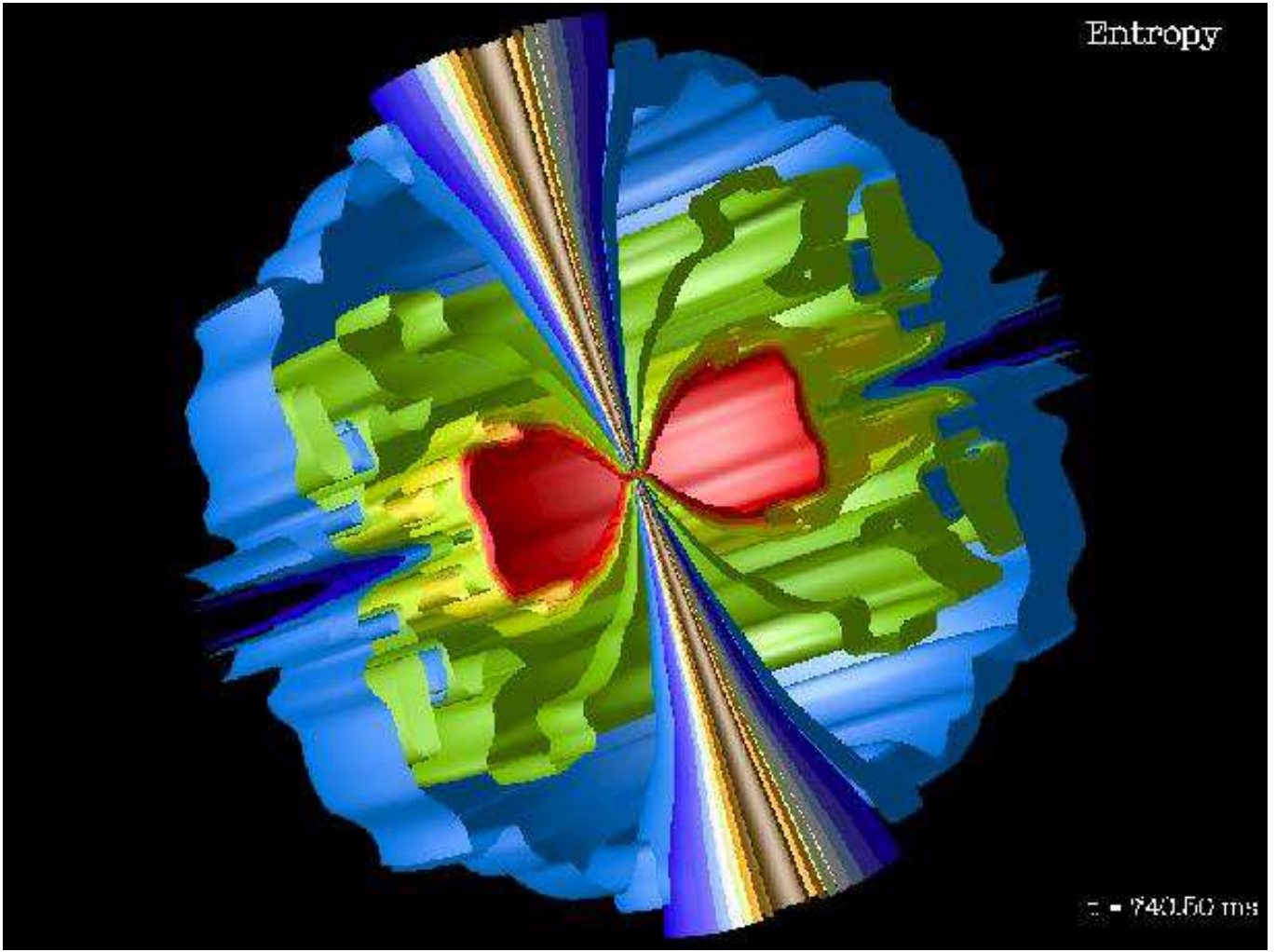, width=1.0\textwidth}
\caption{
A 3D still taken during the explosion of the 1.92-M$_{\odot}$ model of Dessart et al. \cite{dessarta}
that portrays the bipolar neutrino-driven wind that emerges in this model.  The
progenitor is a very rapidly rotating white dwarf. Figure \ref{aburrows_flux}
depicts the corresponding degree of angular anisotropy of the neutrinos that emerge from the
oblate, spinning protoneutron star.  It is this severe anisotropy that translates into
the gross asymmetry of the resulting wind.
}
\label{aburrows_fig_aic1}
\end{figure*}

Recently, Dessart et al. \cite{dessarta} performed a study of the 
accretion-induced collapse (AIC) of white dwarfs into neutron stars,
including the effects of the rapid rotation that must certainly attend accretion by the primary
white dwarf on its way to achieving the Chandrasekhar mass and core collapse.  A major distinction
between AIC and canonical core collapse is the absence in the former of a significant stellar envelope.
Hence, neutrino driving is not inhibited.  The shock generated at bounce moves slowly, but steadily, outwards.
Within 50--100\,ms, the stalled shock breaks out of the white dwarf along the poles.
Figure \ref{aburrows_fig_aic1} depicts the bipolar morphology of a representative blast
following an AIC. The blast is followed, 200-300\,ms after bounce, by a neutrino-driven wind that develops within
the excavated white dwarf, in a cone of $\sim$40$^{\circ}$ opening angle about the pole, with a mass
loss rate of 5-8$\times$10$^{-3}$\,M$_{\odot}$\,s$^{-1}$. The ejecta have an entropy on the order of
20-50\,$k_B$/baryon, and the electron fraction is bimodal, with peaks at 0.25 (due to
the neutrino-driven wind) and 0.5 (due to the original blast and the wind along the pole).
Hence, the AIC of white dwarfs leads to successful explosions with modest energy
$\le 10^{50}$\,erg, thus comparable to the energies obtained through the collapse of
O/Ne/Mg core of stars with $\sim$8-11\,M$_{\odot}$ main sequence mass 
\cite{kitaura,buras2}. This is, however, underenergetic, by a factor 
of about ten, compared with the inferred value for the core collapse 
of more massive progenitors leading to Type IIP supernovae.
Therefore, the neutrino mechanism can successfully power low-energy 
explosions of low-mass progenitors and AICs due to the
limited mantle mass and steeply declining accretion rate.
%
While the total electron-neutrino luminosities are comparable to those observed in standard core-collapse
simulations, the effects of rotation are to reduce the ``$\nu_{\mu}$'' and $\bar{\nu}_e$ luminosities, the
latter by one order of magnitude.
Additionally, the neutron stars resulting from such AIC of white dwarfs are strongly aspherical,
with neutrinospheres having disk-like shapes, akin to a polar-pinched oblate surface.
In the faster rotating model (1.92-M$_{\odot}$) of the Dessart et al. \cite{dessarta} study, 
this configuration results in a strong latitudinal dependence of the neutrino flux, 
enhanced in the polar direction and reduced in the equatorial
direction compared to a non-rotating case (see Fig. \ref{aburrows_flux}).

\begin{figure*}
\epsfig{file=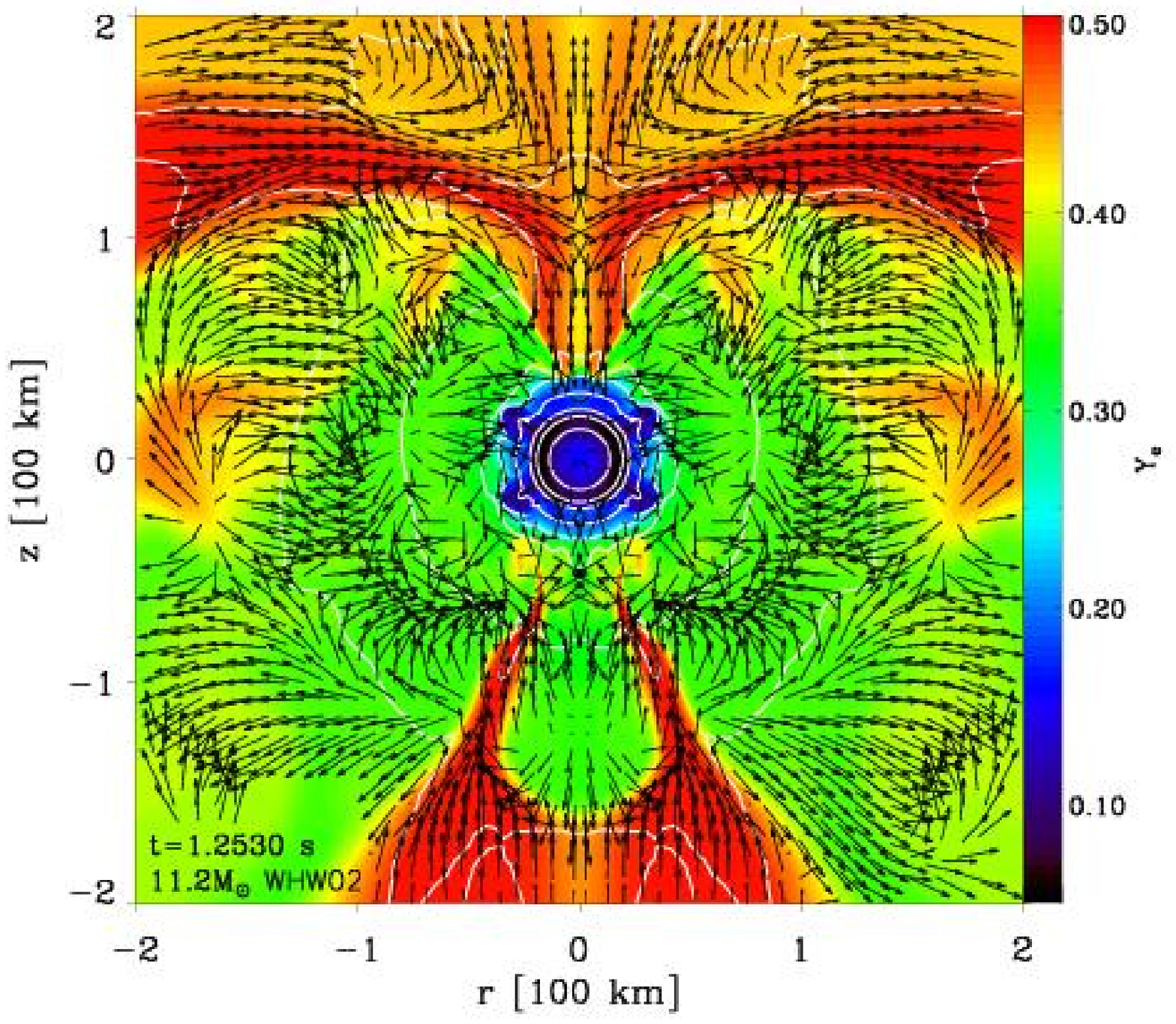, width=0.5\textwidth}  
\epsfig{file=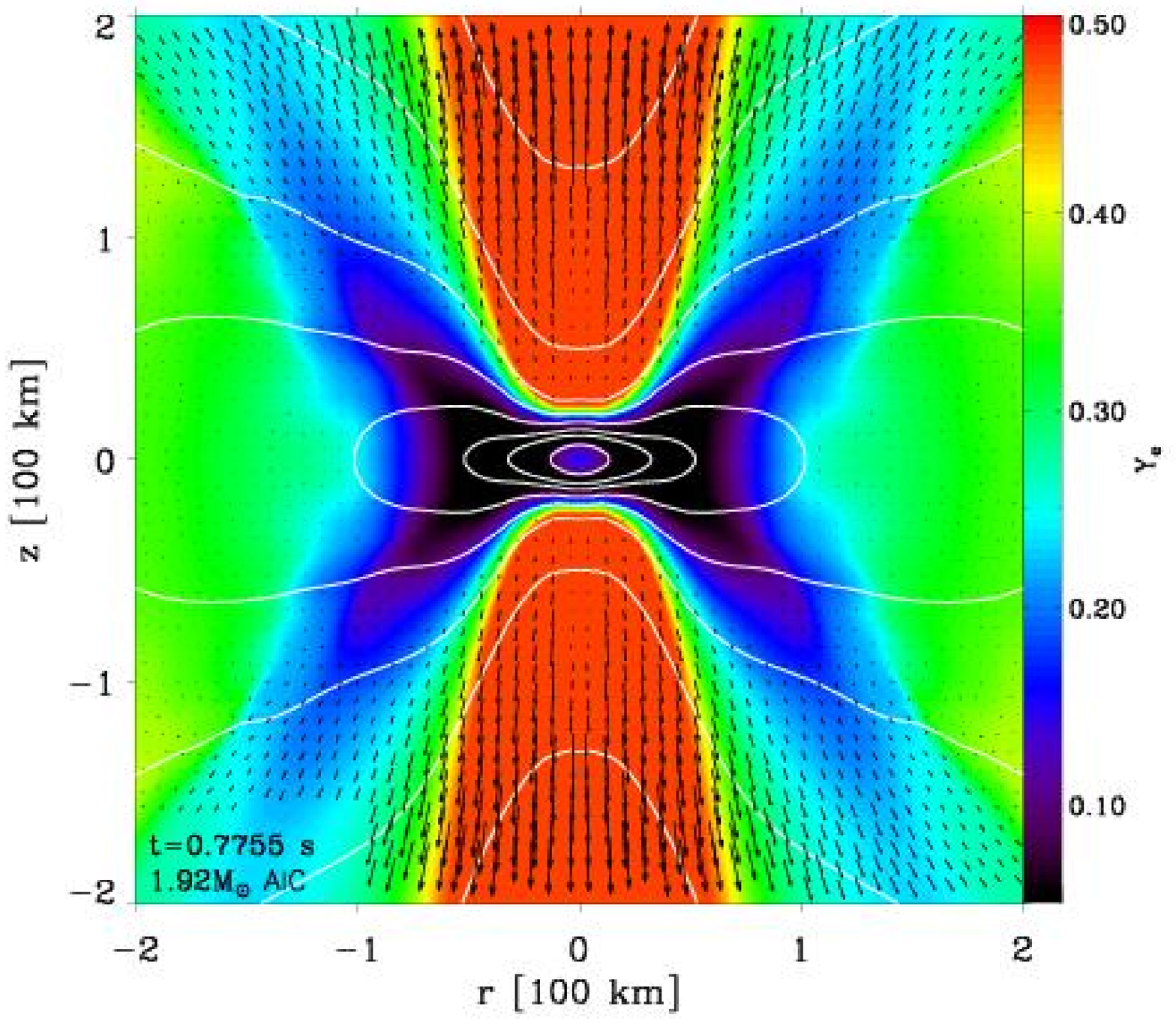, width=0.5\textwidth}  
\caption{
Colormap of the electron fraction for the 11.2-M$_{\odot}$ model of WHW02
at 1.253\,s (left) and for the 1.92-M$_{\odot}$ AIC model of Dessart et al. \cite{dessarta}
at 0.7755\,s (right), both times since bounce.
Velocity vectors are overplotted in black, with a length saturated to 5\% of
the width of the display at $5 \times 10^8$\,km\,s$^{-1}$.
Note the large convective motions in the left panel, occurring between the
shock region and the surface of the protoneutron star. By contrast,
the AIC model exhibits a strong neutrino driven wind along the poles, with little or
no convection within the protoneutron, in part inhibited by the fast rotation of the
corresponding layers.
}
\label{aburrows_ye}
\end{figure*}

The deleptonized region connected to the neutrinosphere has a butterfly, rather than a
spherical shape. Moreover, the neutrino-driven wind originating in the vicinity of the
distorted neutrinosphere sees a lower electron-neutrino flux further away from the poles,
resulting in a latitudinal dependence of the electron fraction of the ejected material.
Figure \ref{aburrows_ye} depicts the Y$_e$ distribution of the ejecta of an AIC explosion,
and compares it to the corresponding distribution for our simulation of the 11.2-M$_{\odot}$ model of WHW02.
Superposed are the 2D $r-z$ velocity vector fields.
%
%
\begin{figure*}
\epsfig{file=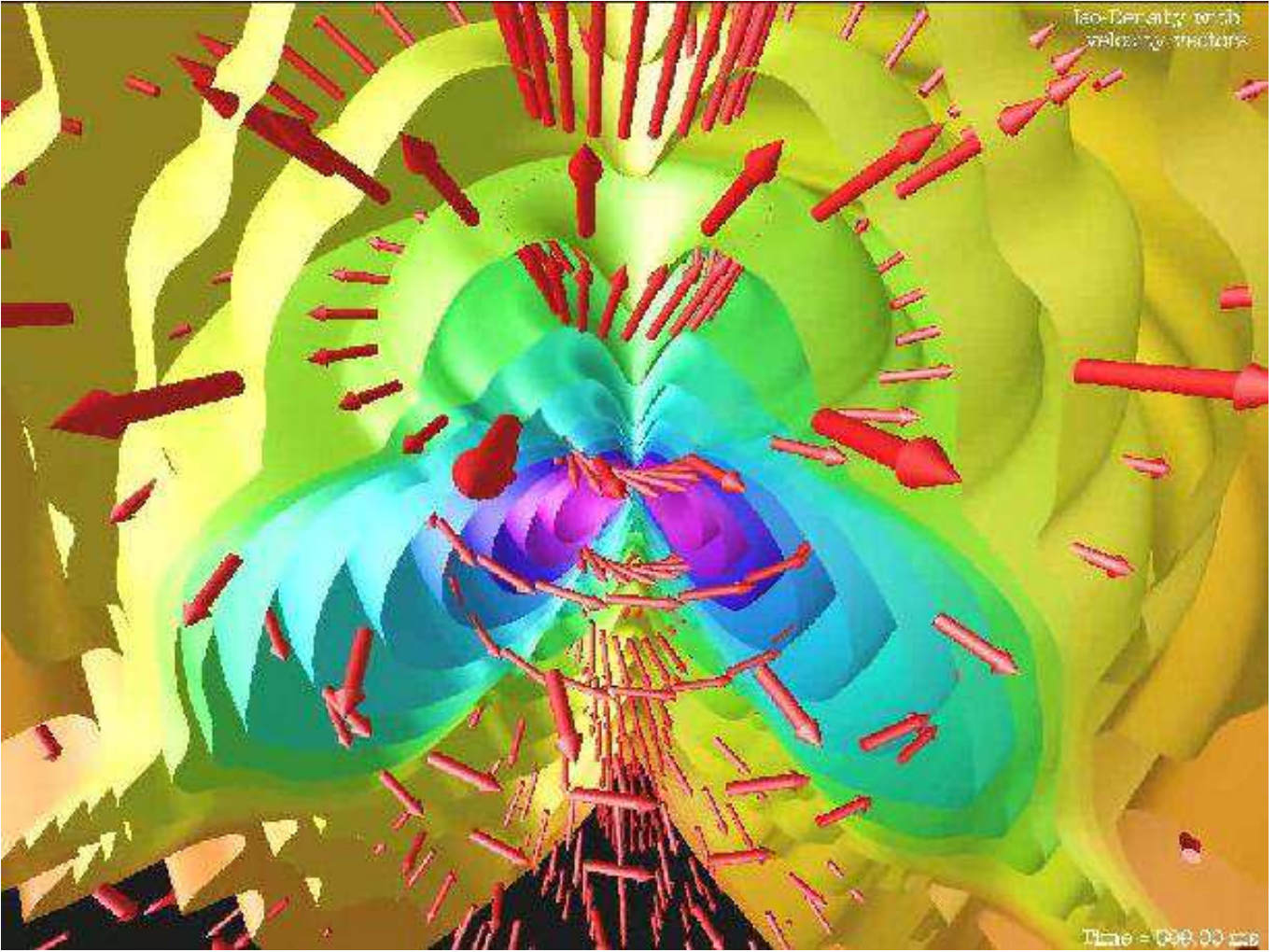, width=1.0\textwidth}
\caption{
A snapshot employing isodensity sheets during the explosive
phase of the simulation of a rapidly rotating white dwarf created in accretion-induced
collapse.  The arrows indicate the direction and magnitude of the velocity vectors,
including the effect of rotation.
This model explodes along the poles, leaving
behind a thick disk of material (Simulation performed and published by Dessart
et al. \cite{dessarta}.)
}
\label{aburrows_fig_aic2}
\end{figure*}
Compared to standard core collapse, our AIC simulations show no perceptible signs of convection associated
with the negative lepton radial-gradient in the protoneutron star, no $\ell=1,2$ oscillations
associated with the vortical-acoustic instability, no late time 
core oscillations, and no sizable neutron star kick.
Moreover, in all models, a quasi-Keplerian 0.1-0.5\,M$_{\odot}$ disk remains in the equatorial region,
that should later be accreted by the neutron star on longer, viscous timescales of many seconds.
Figure \ref{aburrows_fig_aic2} is a 3D rendering, with 3D velocity vectors, of an explosive
stage of an AIC simulation.  The quasi-Keplerian regions and the polar ejecta are clearly 
indicated with the 3D velocity vector field.

\section{Core Convection Effects in the Protoneutron Star}
\label{pns}

Because neutrino heating might still be an important factor in the supernovae mechanism,
an enhancement in the neutrino flux during the first second
following core bounce could certainly facilitate explosion.
Convection in the nascent protoneutron star (PNS) has been invoked as a potential
mechanism for such an increase in the neutrino luminosity.
Neutrino escape at and above the neutrinosphere establishes a negative lepton-gradient, a
situation unstable to convection under the Ledoux criterion.
However, Lattimer \& Mazurek \cite{lm81} challenged the idea of large-scale overturn, noting
the presence of a positive, stabilizing entropy gradient - a residue of the shock birth.

In addition, Mayle \& Wilson \cite{mayle,wm88,wm93} have argued
that regions stable to convection according to the Ledoux criterion could be sites
of doubly-diffusive instabilities, taking the form of so-called neutron (low-$Y_e$ material) fingers.
This idea rests essentially on the assumption that the neutrino-mediated diffusion of heat
occurs on shorter timescales than the neutrino-mediated diffusion of leptons.
By contrast, Bruenn \& Dineva \cite{dineva} and Bruenn, Raley, \& Mezzacappa \cite{bruenn2005} demonstrated
that the neutrino-mediated thermal-diffusion timescale is longer than that of
neutrino-mediated lepton-diffusion, and, thus, that neutron fingers do not obtain.
Because  $\nu_{\mu}$'s and $\nu_{\tau}$'s have a weak thermal coupling
to the material, Bruenn, Raley, \& Mezzacappa \cite{bruenn2005} concluded that lepton-diffusion would occur faster
by means of low-energy ${\bar{\nu}}_{\rm e}$'s and $\nu_{\rm e}$'s.
Applying their transport simulations to snapshots of realistic core-collapse simulations,
they identified the potential for two new types of instabilities within the PNS, referred
to as ``lepto-entropy fingers" and ``lepto-entropy semiconvection."

To study these issues, Dessart et al. \cite{dessartb} have recently performed VULCAN/2D simulations
of core convection in the protoneutron star all the way to the center 
and including lateral neutrino transport.  Our conclusions were threefold: First, 
we do not see any large-scale overturn of the inner PNS material.
Second, we see no evidence of doubly-diffusive instabilities in the PNS, expected to
operate on diffusion timescales of at least a second, but instead observe the
presence of convection, within a radius range of 10-20\,km, operating with a
timescale of a few milliseconds.  Indeed, convection, directly connected 
to the PNS, is found to occur in two distinct regions: between
10 and 20\,km, coincident with the region of negative lepton gradient, and
exterior to the PNS above 50\,km. However, these two regions are separated by an
interface, which shows no sizable outward or inward motion
and efficiently shelters the inner PNS. 

We find that such convective motions do not appreciably enhance the $\nu_e$ neutrino luminosity,
and that they can enhance the $\bar{\nu}_e$ and ``$\nu_{\mu}$" luminosities
by no more than $\sim$15\% and $\sim$30\%, respectively,
during the first post-bounce $\sim$100 ms, after which the optical depth
barrier between the inner convection and the neutrinospheres effectively isolates
one from the other, terminating even this modest enhancement.
PNS convection is thus found to be a secondary feature of the core-collapse
phenomenon, rather than a decisive ingredient for a successful explosion.
Furthermore, the typical timescale associated with such convective transport is of
the order of a few milliseconds, and thus is at least a thousand times faster
than typical growth rates for instabilities associated with neutrino-mediated
thermal and lepton diffusion. Such doubly-diffusive instabilities are, 
therefore, unlikely to play a substantial role in the early critical 
phases of the PNS. Hence, we concluded that inner PNS motions do not bear 
importantly on the potential success of core-collapse supernovae explosions.

\section{Gravitational Radiation from Supernovae}
\label{gr}

In the past, model calculations
estimating supernova signatures have focussed on the
gravitational-wave emission from rotating iron core collapse
and core bounce\cite{ott}.
Recent results from stellar evolutionary
calculations \cite{heger05,hirschi,hirschi2} and neutron star birth spin
estimates \cite{ott06b} indicate that presupernova
stellar iron cores may rotate much more slowly
than previously assumed and that the asphericity during collapse
and bounce due to rotation is not generally great enough to produce a sizable
time-varying, wave-emitting mass-quadrupole moment. In fact, gravitational
radiation from large-scale post-bounce convection and anisotropic neutrino
emission are likely to exceed the bounce signal of such slowly rotating
supernova cores.

The first-generation LIGO-class detectors are now operating at design
sensitivity and an international network of observatories, including
LIGO, GEO600, VIRGO and TAMA, is on-line.
Gravitational waves detected from a supernova can provide
us with ``live'' dynamical
information from the supernova core, complementing the supernova neutrino pulse
as the only other immediate carriers of information from deep inside the star.
Using signal-processing techniques operating on a large
set of theoretical templates, it will be possible to extract
supernova physics from a sufficiently strong signal.

The core-oscillation/acoustic mechanism introduces the intriguing possibility 
that strong gravitational waves can arise from its quadrupole
spatial components and, as a result, Ott et al. \cite{ott06a} obtain
new estimates for the gravitational-wave signature of core-collapse
supernovae. When the $\ell = 1$ g-mode grows strong, it starts
transferring energy through nonlinear effects to the harmonic at twice its frequency.  This harmonic has
some $\ell = 2$ spatial character.  In addition, in some of our models, particularly
the more massive ones, fundamental $\ell = 2$ g-modes themselves are excited.  
As a result, Ott et al. \cite{ott06a} found that the gravitational waves from the quadrupole components
of the core oscillations, if they occur, can dominate the total gravitational wave signature
in duration, maximum strain, and total energy emission by one to
several orders of magnitude.  They have also discovered
an approximate progenitor dependence: more massive iron
cores may experience higher frequency, higher
amplitude oscillations, and, hence, more energetic
gravitational-wave emission.  Note that the acoustic model predicts a delay between
core bounce and the onset of the significant phase of gravitational wave emission.
It also suggests that during this phase the frequency spectrum of the waves is much more narrow
than in previous theories.  These are signatures of and tests for the core oscillation model.

\section{Numerical Issues}
\label{num_issues}

\begin{figure*}
\epsfig{file=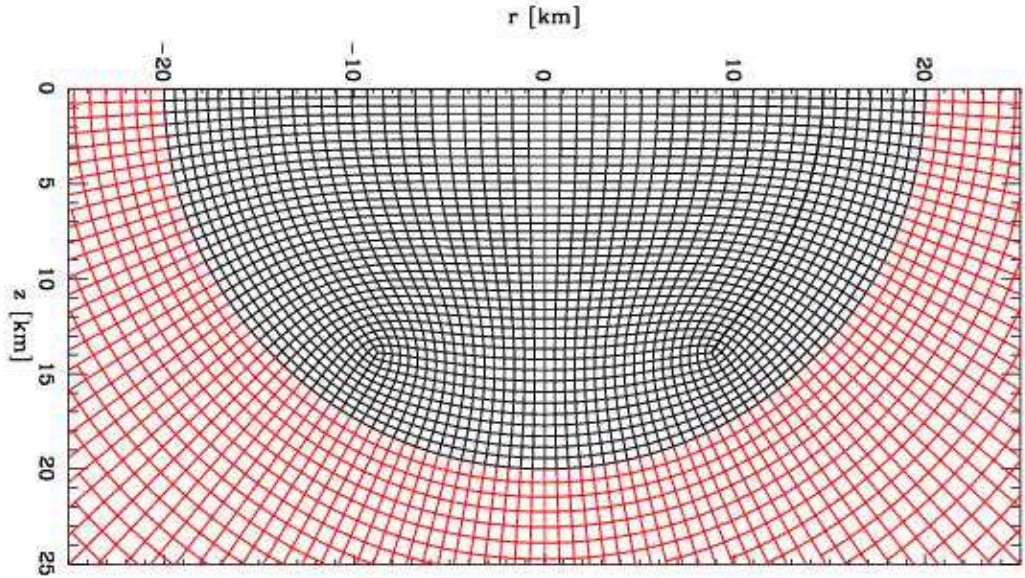, width=1.0\textwidth, angle=-90}
\vspace{-2.5in}
\caption{Illustration of the VULCAN/2D unstructured grid in the central region where
it transitions from spherical-polar (red) to Cartesian (black).
The spherical-polar region extends from the transition radius at 20\,km out to
a maximum radius, with at least 141 logarithmically-spaced points.
A relaxation method is used to construct the node positions in the inner region,
starting with 40 equally-spaced points.
}
\label{aburrows_grid}
\end{figure*}

For these simulations, we have used the two-dimensional
multi-group, multi-neutrino-species, flux-limited diffusion (MGFLD) variant of the
code VULCAN/2D \cite{bur06,bur06b,dessarta,dessartb,ott06b,ott06a,livne04,walder}.
This is currently the only extant 2D, multi-group code that allows core translational 
motion by introducing a Cartesian-like grid in the inner core and, hence, that
is capable of investigating the core-oscillation/acoustic mechanism. 
Figure \ref{aburrows_grid} gives an example of such a grid.  Note the transition
``nodes" at the juncture between the inner and outer griddings.
%
VULCAN/2D is still the only extant supernova code to perform 2D (not ``ray-by-ray") 
multi-group transport with general gravity all the way to the center.  Due to the
finite-difference character of 2D codes that employ spherical coordinates 
all the way to the center, to the singularity in those coordinates at that center,
and to the reflecting boundary condition frequently imposed at this center,
spherical-coordinate codes are likely to inhibit core translational motions
artificially and, hence, to inhibit the $\ell = 1$ g-modes that are central to
the mechanism we have identified. Be that as it may,
there are many caveats to our study the reader should keep in mind:  1) Our
calculations are Newtonian and not general-relativistic; 2) We
employ an approximate (MGFLD, not multi-angle) multi-group transport 
algorithm in the neutrino sector; 3) Numerical errors
are bound to have accumulated due to the need to calculate for $\sim$1,000,000 timesteps
for each progenitor; 4) The initial seed perturbations are unknown (and unknowable?);
5) We are performing the calculations in 2D, not 3D; 
6) The Doppler-shift terms in the transport equation
have been dropped. While these velocity-dependent terms are 
very different in the laboratory-frame formulation \cite{hubeny}
we have adopted than in the comoving frame formulation of
Buras et al. \cite{buras2006}, they should nevertheless be incorporated;
7) We have used only 16 energy groups; using more ($\ge 20$) is preferred
\cite{Tod1}; 8) The transition in the grid from Cartesian-like
to spherical coordinates introduces slight artifacts 
in the flow at the associated topological junctures. The 
spatial resolution in the center is good, but can be improved on 
the outside exterior to $\sim$200 km; and 9) The opacities employed
are sophisticated\cite{bur04f}, but the neutrino-matter correlation effects
at higher densities need a second look.  

Importantly, the flow is fundamentally chaotic and a precise mapping between initial
configuration and final outcome is not possible.  This multi-dimensional radiation/hydrodynamical
problem is quintessentially meteorological in character.  Nevertheless, 
ours are the first calculations to explore
the novel core oscillation/acoustic mechanism and to venture into the 
late-time behavior of multi-D core collapse with multi-D core motions
and multi-D/multi-group transport.
What the actual and relative contributions of sound and neutrinos are 
to the supernova phenomenon as a function of progenitor   
remains to be determined and will require even more sophisticated
numerical tools than we have applied to date to reach a definitive answer.
It may be that the neutrino mechanism still obtains in the generic
case and that the resulting earlier explosion aborts the acoustic mechanism.

\section{BETHE: A New Multi-Dimensional Radiation Hydrodynamic Code}
\label{bethe_code}

Many phenomena in the Universe must be addressed
using the tools of radiation transport, radiative
transfer, and radiation hydrodynamics to achieve a
theoretical understanding of their character.  As this monograph 
demonstrates, supernovae are among them and for them neutrinos 
are the radiation.  However, whether the radiation is 
neutrinos or photons, time-dependent techniques to address 
radiation transport and the coupling of radiation with matter are
of central concern to the theorist whose goal is explaining 
astrophysical phenomena. Moreover, spherically-symmetric algorithms
are often not sufficient and multi-dimensional approaches are called for.
These are not easy, not only to formulate, but to implement.
Nevertheless, multi-D radiation transport is emerging as
a necessary tool in the theorist's toolbox and computers
to evolve the associated equations are becoming available
to a wider cohort of researchers.   

In that spirit, Hubeny \& Burrows \cite{hubeny} have derived and are starting to implement
a new {\it mixed-frame}, multi-D, velocity-dependent 
radiation code that is a departure from the approaches
employed to date in supernova theory and in astrophysics.
The mixed-frame approach, in which the radiation quantities are
defined in the laboratory (Eulerian) frame and the matter and coupling
quantities are defined in the comoving frame, has largely been
neglected by the radiation transport and atmospheres communities
because of their focus on line transfer.  Their need to
include narrow spectral lines and to handle Doppler shifts into
and out of those lines necessitates many spectral bins and a large 
number of angular bins to ensure the lines are resolved on the computational
grid.  As a result, most dynamic atmosphere and radiation studies are done 
using the comoving (Lagrangian) equations of radiation transport.
However, since many radiation-hydrodynamic problems do
not require an exquisite treatment of spectral line transport, but
a good treatment of continuum transport, the mixed-frame methodology is perfectly-well
suited to many astrophysical problems. Its virtues vis \`a vis the comoving-frame approach
to the solution of the Boltzmann equation and its related moment
equations are numerous: 1) Even in one-dimension, instead of requiring
$\sim$20 velocity-dependent terms on the left-hand(operator) side of the
Boltzmann/transport equation, many of which involve spatial velocity
derivatives, there are no such terms on the left-hand-side in the mixed frame approach
and only {\it one} grouped linear term (to $O(v/c)$) on the right-hand(source)
side;  2) There are no terms with derivatives
of the velocity. Therefore, the characteristics of the associated
transport equation are all straight lines.  Furthermore, there
is no need for the monotonicity in the velocity field required
by some implicit solvers; and 3) The mixed-frame method is easily
generalized to two and three dimensions, and the associated solvers
are straightforward (though more expensive) extensions of those employed in
1D.  Note that much has been made of the importance of velocity-dependent
terms in the transport equations for the calculation
of the neutrino energy deposited in the ``net-gain"
\cite{wilson85} region.  However, the mixed-frame approach provides the
clearest perspective from which to understand the physics 
of this effect and to see it as simple Doppler shifts of the intensity 
and neutrino energies. The comoving frame approach obscures this 
straightforward fact. 

We have christened our new radiation-hydrodynamics code employing the
mixed-frame philosophy BETHE\footnote{{\bf B}asic (2-Dimensional), 
{\bf E}xplicit/Implicit, {\bf T}ransport and {\bf H}ydrodynamics
{\bf E}xplosion (Code)}.   Our formulation is multi-group
and multi-angle and includes anisotropic scattering,
frequency(energy)-dependent scattering and absorption,
complete velocity dependence to order $v/c$, rotation, and
energy redistribution due to inelastic scattering. Hence,
the ``2D" realization is actually ``6 1/2"-dimensional.
The effects of radiation viscosity are automatically incorporated
and we have developed Accelerated-Lambda-Iteration, Krylov subspace (GMRES),
Discontinuous-Finite-Element, and Feautrier numerical methods for solving
the equations.  Such new algorithms are necessary to improve sufficiently
upon what has been done before and explore the viability of both
the neutrino and the acoustic mechanisms. Despite the new ideas summarized
in this paper, we would not claim that the supernova problem is close to being solved.  
Far from it.  There are always numerical issues that cloud the 
interpretations of numerical results and constant evolution
of technique is necessary to help validate or refute theory.  However, we hope that
the new code, BETHE, will prove to be a step forward in the ongoing quest to understand
one of Nature's most vexing theoretical puzzles and that Hans, who did so much to advance
and promote supernova research, would have looked upon our all-too-immodest 
and inadequate gesture to honor him with an approving smile.

\acknowledgments

We acknowledge partial support for this work
from the Scientific Discovery through Advanced Computing
(SciDAC) program of the DOE, grant number DE-FC02-01ER41184,
and from the NSF under grant AST-0504947.  Furthermore, we 
thank the Joint Institute for Nuclear Astrophysics (JINA) for 
a subaward to the University of Arizona.


\begin{thebibliography}{999}



\bibitem{bethe}
H. Bethe, J.R. Wilson, \apj {\bf 295} (1985) 14.

\bibitem{buras2006} R. Buras, M. Rampp, H.-Th. Janka, K. Kifonidis, 
\aa {\bf 447} (2006a) 1049.

\bibitem{buras2} R. Buras, H.-Th. Janka, M. Rampp, 
K. Kifonidis, accepted to \aap (2006b).  

\bibitem{lieben2001}
M. Liebend\"{o}rfer, A. Mezzacappa, F.-K. Thielemann, O.E.B. Messer,
W.~R. Hix, S.W. Bruenn, \prd {\bf 63} (2001) 103004.

\bibitem{wm88} J.R. Wilson, R. Mayle, {\em Phys. Rep.} {\bf 163} (1988) 63.

\bibitem{wm93} J.R. Wilson, R. Mayle, {\em Phys. Rep.} {\bf 227} (1993) 97.

\bibitem{mayle} R. Mayle, J.R. Wilson, \apj {\bf 334} (1988) 909. 

\bibitem{herant}
M. Herant, W. Benz, W.R. Hix, C.L. Fryer, S.A. Colgate, \apj {\bf 435} (1994) 339.

\bibitem{bhf}
A. Burrows, J. Hayes, B.A. Fryxell, \apj {\bf 450} (1995) 830.

\bibitem{muller96} H.-Th. Janka, E. M\"{u}ller, \aa {\bf 306} (1996) 167.

\bibitem{fryer2002} C.L. Fryer, M. Warren, \apj {\bf 574} (2002) L65.

\bibitem{fryer2004} C.L. Fryer, M. Warren, \apj {\bf 601} (2004) 391.

\bibitem{kitaura} F.S. Kitaura, 
H.-Th. Janka, W. Hillebrandt, \aa {\bf 450} (2006) 345.

\bibitem{nomoto} K. Nomoto, M. Hashimoto, {\em Phys. Repts.} {\bf 163} (1988) 13.

\bibitem{woosley02}
S.E. Woosley, A. Heger, T.A. Weaver, {\em Rev. Mod. Phys.} {\bf 74} (2002) 1015.

\bibitem{blond03} J.M. Blondin, 
A. Mezzacappa, C. DeMarino, \apj {\bf 584} (2003) 971.

\bibitem{blond06} J.M. Blondin,
A. Mezzacappa, \apj {\bf 642} (2006) 401. 

\bibitem{fogt00} T. Foglizzo, M. Tagger, \aa {\bf 363} (2000) 174.

\bibitem{fog01} T. Foglizzo, \aa {\bf 368} (2001a) 311.

\bibitem{fog02} T. Foglizzo, \aa {\bf 392} (2001b) 353.

\bibitem{fog5a}  T. Foglizzo, P. Galletti, M. Ruffert, \aa {\bf 435} (2005) 397.

\bibitem{bur06} A. Burrows, E. Livne, L. Dessart, C.D. Ott, J. Murphy,
\apj {\bf 640} (2006a) 878.

\bibitem{bur06b} A. Burrows, E. Livne, L. Dessart, C.D. Ott, J. Murphy,
submitted to \apj (2006b).

\bibitem{fog06} T. Foglizzo, P. Galletti, L. Scheck, 
H.-Th. Janka, submitted to \apj (2006). 

\bibitem{bg93} A. Burrows, J. Goshy, \apj {\bf 416} (1993) 75.

\bibitem{scheck}
L. Scheck, T. Plewa, H.-Th. Janka, K. Kifonidis, E. M\"uller, \prl {\bf 92} (2004) 011103.

\bibitem{scheck2}
L. Scheck, K. Kifonidis, H.-Th. Janka, E. M\"uller, accepted to \aap (2006).

\bibitem{wang2}
L. Wang, D. Baade, P. H\"oflich, J.C. Wheeler, \apj {\bf 592} (2003) 457. 

\bibitem{leonard} D.C. Leonard, et al.,  {\em Nature} {\bf 440} (2006) 505.

\bibitem{cordes} J.H. Taylor, J.M. Cordes, \apj {\bf 411} (1993) 674.

\bibitem{lyne} A. Lyne, D.R. Lorimer, {\em Nature} {\bf 369} (1994) 127.

\bibitem{dessarta} L. Dessart, A. Burrows, C.D. Ott, E. Livne, 
S.-Y. Yoon, N. Langer, \apj {\bf 644} (2006a) 1063.

\bibitem{lm81} J.M. Lattimer, T.J. Mazurek, \apj {\bf 246} (1981) 955.

\bibitem{dineva} S.W. Bruenn, T. Dineva, \apj {\bf 458} (1996) L71.

\bibitem{bruenn2005} S.W. Bruenn, E.A. Raley,
A. Mezzacappa, astro-ph/0404099 (2005).

\bibitem{dessartb} L. Dessart, A. Burrows, E. Livne, C.D. Ott, 
\apj {\bf 645} (2006b) 534.  

\bibitem{ott} C.D. Ott, A. Burrows, E. Livne, R. Walder, 
\apj {\bf 600} (2004) 834.

\bibitem{heger05} A. Heger, S.E. Woosley,
H. Spruit, \apj {\bf 626} (2005) 350. 

\bibitem{hirschi} R. Hirschi, G. Meynet, A. Maeder, \aap {\bf 425} (2004) 649. 

\bibitem{hirschi2} R. Hirschi, G. Meynet, A. Maeder, {\em Nucl. Phys. A} {\bf 758} (2005) 234.

\bibitem{ott06b} C.D. Ott, A. Burrows, L. Dessart, E. Livne, 
\apj~Suppl. {\bf 164} (2006a) 130.  

\bibitem{ott06a} C.D. Ott, A. Burrows, L. Dessart, E. Livne, 
\prl {\bf 96} (2006b) 201102.  

\bibitem{livne04} E. Livne, A. Burrows, R. Walder, 
T.A. Thompson, I. Lichtenstadt, \apj {\bf 609} (2004) 277.

\bibitem{walder} R. Walder, A. Burrows, C.D. Ott,
E. Livne, I. Lichtenstadt, M. Jarrah, \apj {\bf 626} (2005) 317.

\bibitem{hubeny} I. Hubeny, A. Burrows, submitted to {\it Astrophys.~J.} (2006).

\bibitem{Tod1} T.A. Thompson, 
A. Burrows, P.A. Pinto, \apj {\bf 592} (2003) 434.

\bibitem{bur04f} A. Burrows, T.A. Thompson, 2004, ``Neutrino-Matter Interaction
Rates in Supernovae: The Essential Microphysics of Core Collapse,"
in {\it Core Collapse of Massive Stars},
C. Fryer Ed., Kluwer Academic Press, 2004, p. 133.

\bibitem{wilson85} J.R. Wilson, in {\em Numerical Astrophysics},
J. Centrella, J. M. LeBlanc, R. L. Bowers, Eds., Jones \& Bartlett, Boston, 1985, p. 422.


\end{thebibliography}
\end{document}